\documentclass[iop]{emulateapj}
\usepackage{natbib}
\usepackage{threeparttable}
\usepackage{subfigure}
\usepackage{amssymb,amsmath}
\usepackage{hyperref}
\usepackage{graphicx}
\usepackage{epstopdf}
\usepackage{natbib}
\usepackage{rotating} 

\newcommand{\EQ}{\begin{equation}}
\newcommand{\EN}{\end{equation}}
\newcommand{\EQA}{\begin{eqnarray}}
\newcommand{\ENA}{\end{eqnarray}}
\newcommand{\bfu}{\mathbf{u}}
\newcommand{\bfx}{\mbox{\boldmath $x$} {}}
\newcommand{\bfB}{\mathbf{B}}
\newcommand{\bfl}{\mbox{\boldmath $l$} {}}
\newcommand{\uu}{\mbox{\boldmath $u$} {}}

\newcommand{\BB}{\mbox{\boldmath $B$} {}}
\def\half{{\textstyle{1\over2}}}

\begin{document}
\pagestyle{plain}
\pagenumbering{arabic}

\title{Mixing in Magnetized Turbulent Media}   
\author{Sharanya Sur\altaffilmark{1}, Liubin Pan\altaffilmark{2} \& Evan Scannapieco\altaffilmark{1}}
\altaffiltext{1}{School of Earth and Space Exploration, Arizona State University, 
PO Box 876004, Tempe - 85287, USA}
\altaffiltext{2}{Harvard-Smithsonian Center for Astrophysics, 60 Garden St., Cambridge MA 02138, USA}
\email{sharanya.sur@asu.edu, lpan@cfa.harvard.edu, evan.scannapieco@asu.edu}

%-------------------------------------------------------------------------------------------------------------------------------

\begin{abstract}

Turbulent motions are essential to the mixing of entrained fluids and are 
also capable of amplifying weak initial magnetic fields by small-scale dynamo 
action. Here we perform a systematic study of turbulent mixing in magnetized 
media, using three-dimensional magnetohydrodynamic simulations that include 
a scalar concentration field. We focus on how mixing depends on the magnetic 
Prandtl number, Pm, from 1 to 4 and the Mach number, ${\mathcal M},$ from 0.3 to 
2.4. For all subsonic flows, we find that the velocity power spectrum has a $k^{-5/3}$ 
slope in the early, kinematic phase, but steepens due to magnetic back reactions 
as the field saturates. The scalar power spectrum, on the other hand, flattens 
compared to $k^{-5/3}$ at late times, consistent with the Obukohov-Corrsin picture 
of mixing as a cascade process. At higher Mach numbers, the velocity power spectrum 
also steepens due to the presence of shocks, and the scalar power spectrum again 
flattens accordingly. Scalar structures are more intermittent than velocity structures 
in subsonic turbulence while for supersonic turbulence, velocity structures appear
more intermittent than the scalars only in the kinematic phase. Independent of the 
Mach number of the flow, scalar structures are arranged in sheets in both the kinematic 
and saturated phases of the magnetic field evolution. For subsonic turbulence, scalar 
dissipation is hindered in the strong magnetic field regions, probably due to Lorentz 
forces suppressing the buildup of scalar gradients, while for supersonic turbulence, 
scalar dissipation increases monotonically with increasing magnetic field strength. 
At all Mach numbers, mixing is significantly slowed by the presence of dynamically-important 
small-scale magnetic fields, implying that mixing in the interstellar medium and in galaxy 
clusters is less efficient than modeled in hydrodynamic simulations.

\end{abstract}


%-------------------------------------------------------------------------------------------------------------------------------

\section{Introduction}\label{intro}

Turbulent mixing is an important topic with many applications \citep{D05}, from  
atmospheric research \citep[e.g.,][]{J01}, to combustion \citep[e.g.,][]{P06}, to astrophysics.   
Within and around our galaxy, for example, mixing in turbulent flows is essential to 
interpret a wide variety observations including the metallicity dispersion in open clusters 
\citep{FB92,Q02,Desilva+06}, the abundance of scatter along different lines of sight in the 
interstellar medium (ISM) \cite{Cart+06}, and the cluster to cluster metallicity scatter \citep{T+97}. 
At moderate redshifts, mixing is a key process in the enrichment of the intergalactic medium
\citep[e.g,][]{Schaye+03,Pichon+03, Pieri+06,Scan+06,Becker+09} and the distribution of metals 
in galaxy clusters \citep{reb+05,david+08,scan+09}. At the highest redshifts, mixing determines 
the pollution of primordial gas by the first generation of stars in early galaxies \citep{PS07,PSSc12, 
PSS13} and the transition from Population III to Population II star formation \citep{SSF03}. 

Over the course of the last decades, advances in numerical modeling of turbulent flows have 
broadened our understanding of turbulent mixing. This has led to a confirmation of the physical 
picture of turbulent mixing, often referred to as the `Obukohov-Corrsin' (OC) phenomenology 
\citep{SS00} in which scalar mixing is described as a cascade process caused by the stretching 
of the concentration field by the velocity field. The scalar field advected by the turbulent flow thus 
exhibits a complex, chaotically evolving structure over a broad range of spatial and temporal scales. 
Random stretching by the velocity field leads to the production of scalar structures of progressively 
smaller size, down to the scale at which molecular diffusion eventually homogenizes the scalar 
fluctuations. This description has so far been also confirmed numerically for both incompressible 
and compressible hydrodynamic turbulence \citep{SS00,PS10, PS11, PSS13}. 

The turbulent motions responsible for mixing also amplify and maintain magnetic fields 
in astrophysical systems by the dynamo mechanism wherein kinetic energy in turbulent 
fluid motions is tapped to amplify the magnetic energy \citep[see][for a review]{BSS12}. 
The basic underlying mechanism responsible for this dynamo action is the random stretching 
and folding of the field lines by the turbulent eddies.  This leads to initial exponential 
amplification of the magnetic field, followed by an intermediate stage of linear growth as 
fields on progressively larger scales reach dynamically important strengths, and finally by 
a saturated phase as fields on all length scales become large enough to resist further stretching. 

There are two main types of astrophysical dynamos. Large-scale dynamos generate magnetic 
fields on scales larger than the energy-carrying scale of turbulence, whereas small-scale 
dynamos generate magnetic fields on scales comparable to the scale of turbulence. 
While large-scale magnetic fields require special conditions like the presence of helical turbulence
\citep{KR80, RSS88}, small-scale dynamos are generic in any random turbulent flow in which 
the magnetic Reynolds number ${\rm Rm}$, exceeds a critical value ${\rm R_{\rm cr}} \approx 35-100$ 
\citep{BS05}. Such dynamos are capable of amplifying fields on the eddy turnover time scale, which 
is much shorter than the overall age for many astrophysical systems, including galaxies and galaxy 
clusters. This implies that the small-scale dynamo will be crucial for magnetic field generation in 
many astrophysical objects including systems in which conditions for large-scale dynamo operation 
is unlikely.  For example, recent high-resolution numerical simulations have also hinted at the possible 
presence of magnetic fields in the first stars \citep{Sur+10,Fed+11a,Turk+12,Bromm13,Glover13, Latif13}, 
amplified by small-scale dynamo action from seed fields obtained by the Biermann battery 
mechanism \citep{B50} in the early Universe.  

Given the omnipresent nature of astrophysical magnetic fields, it is important to study how mixing 
occurs in a magnetized medium in which the small-scale dynamo is operating. As both mixing and 
the small-scale dynamo are driven by turbulence, this situation will arise naturally in astrophysical 
systems as diverse as the metal-rich interstellar medium \citep{dea+02}, the inhomogeneously-enriched 
intracluster medium within galaxy clusters, \citep{ROh+10,Sakuma+11,Churazov+12}, and the gas 
collapsing onto the very first galaxies, into which metals are being mixed for the first time \citep{PSS13}. 
Here, we address this issue using non-ideal magnetohydrodynamic (MHD) numerical simulations in 
which turbulence is randomly forced in the presence of initially weak magnetic fields. Furthermore, we 
characterize pollutants such as metals as a scalar field satisfying the advection-diffusion equation 
and allow for continuous injection of pollutants on the driving scale of turbulence. 

The structure of the paper is as follows. In Section~\ref{init}, we describe the numerical method and 
the simulation setup used in our study. The results are presented in Section~\ref{results}, where we 
discuss the time evolution our simulations, present a qualitative and quantitative discussion of the 
structure of the density, velocity, magnetic and the scalar concentration fields, and evaluate the impact 
of the small-scale dynamo on the overall mixing timescale.  A summary of our results and conclusions 
is given in Section~\ref{summ}. 
%-------------------------------------------------------------------------------------------------------------------------------

\section{Numerical Method and Initial conditions}\label{init}

To study turbulent mixing in the presence of evolving magnetic fields, we simulated 
continuously-driven, 
three-dimensional turbulence in a periodic box with an initially weak magnetic field using the Eulerian 
grid code FLASHv4\footnote{http://flash.uchicago.edu/site/flashcode/} \citep{Fryxell00}. We solved 
the full set of non-ideal MHD equations given below on a uniform grid with a resolution of $512^{3}$ 
points for a domain of unit size with periodic boundary conditions. \\
\begin{eqnarray}
\partial_t \rho + \nabla\cdot\left(\rho \bfu\right)&=&0 ,\\
\rho \left[
\partial_t \bfu + (\bfu \cdot \nabla) \bfu \right] 
&=& 
(\bfB \cdot \nabla ) \bfB
- \nabla p'  + \nabla\cdot\left(2\nu\rho\boldsymbol{\mathcal{S}}\right) \nonumber \\
&+& \rho{\bf F}, \\
\partial_t E + \nabla\cdot\left[E \bfu \right]  & = & - 
 \nabla\cdot \left( p' \bfu \right)  + \nabla\cdot \left[ \left(\bfB\cdot\bfu\right)\bfB \right] \nonumber \\
 &+& \nabla\cdot \left[ \bfB\times\left(\eta\nabla\times\bfB\right) +
2\nu\rho\bfu\cdot\boldsymbol{\mathcal{S}}\right],\\
\partial_t \bfB  + (\bfu \cdot \nabla) \bfB & = &( \bfB \cdot \nabla) \bfu -\bfB(\nabla\cdot\bfu) + \eta\nabla^2\bfB, \\
\nabla\cdot\bfB  &=  & 0, 
\label{eqnmhd}
\end{eqnarray}
where $\rho$, $\bfu$, $p'=p+ \half\left|\bfB\right|^2$, $\bfB$, and $E=\rho \epsilon_\mathrm{int} + 
\half\rho\left|\bfu\right|^2 + \half\left|\bfB\right|^2$ denote density, velocity, pressure (thermal and magnetic), 
magnetic field, and total energy density (internal, kinetic, and magnetic). The first term on the right hand 
side of eq.~(2) accounts for the magnetic tension due to Lorentz forces, and the
other component of the Lorentz force, the magnetic pressure, is included in the second term. 
In this equation, the last term is the artificial 
driving term for turbulence, and 
viscous and resistive interactions are included in eqs.\ (3-5) via the traceless rate of strain tensor, 
$\mathcal{S}_{ij}=(1/2)(\partial_i u_j+\partial_j u_i)-(1/3)\delta_{ij}\partial_{k}u_{k}$, and $\nu$ and $\eta$ 
are the kinematic viscosity and the magnetic resistivity respectively. The first term on the right hand side 
in eq.~(3) is the $p dV$ work in the presence of magnetic fields while the second term corresponds to 
a component derived from the Lorentz force.
The stretching of the magnetic field lines by a turbulent flow corresponds to the first term on the right hand 
side of eqn.~(4), while the second term denotes compression of the field lines. 
The last term accounts for magnetic field dissipation. 
Our simulations employed the unsplit staggered mesh algorithm in FLASHv4 with a constrained transport
scheme to maintain $\nabla\cdot\BB=0$ to machine precision \citep{LD09, Lee13} and the HLLD Riemann 
solver \citep{MK05}.

In addition to these equations, pollutants are characterized as a scalar field, $C,$ which obeys the 
advection-diffusion equation
\EQ
{\partial_t}\left({\rho C}\right) + \nabla\cdot\left[\rho C\bfu - \rho\kappa \nabla C\right] = \rho{\bf T}, 
\label{Ceq}
\EN
where $\kappa$ is the scalar diffusion coefficient and ${\bf T}$ is the random forcing term for driving 
the scalars. We included these scalar diffusion and forcing terms in a separate module in 
FLASHv4. As described in detail in \cite{PS10}, an equation for the evolution of density-weighted 
variance can be derived from eq.\ (\ref{Ceq}) with the help of the continuity equation, yielding 
\begin{eqnarray}
\partial_t  \langle \tilde{\rho} C^2 \rangle + \partial_i \langle \tilde{\rho} C^2 u_i \rangle 
&=& 2 \langle \partial_i (\tilde{\rho} \kappa C \partial_i C) \rangle  \nonumber \\
& & -  2  \langle \tilde{\rho} \kappa (\partial_i C)^2 \rangle  
+  2 \langle \tilde{\rho} T C \rangle.  
\label{eq:Cvariance}
\end{eqnarray}
where the density-weighting factor $\tilde{\rho}$ is the ratio of $\rho$ to the average flow density,  
$\bar{\rho}$, in the flow, and $\langle \cdot \cdot \cdot \rangle$ denotes an ensemble average, which 
is equal to the average over the flow domain for statistically homogeneous flows, as in our simulations.

The second term on the left hand side of this equation is an advection term, which corresponds to 
the spatial transport of concentration fluctuations between different regions.  In the statistically homogeneous 
case, as in our simulations, this term vanishes, and in its absence, eq.\ (\ref{eq:Cvariance}) does not 
have an explicit dependence on the velocity field. This implies that the velocity field does not truly mix.  
In fact, this also holds true in an inhomogeneous flow. While the advection term is non-zero in this case, 
it is a surface term, and thus its integral over the entire flow domain is zero, which means that advection 
conserves the global scalar variance.

The first two terms on the right hand side of the above equation originate from the scalar diffusion 
term in eq.~(\ref{Ceq}). The first of these again vanishes in the homogeneous case and leads to a 
conservation of  the global scalar variance in the inhomogeneous case. The term 
$\bar{\epsilon}_{\rm c} \equiv -2\langle \tilde{\rho} \kappa (\partial_i C)^2 \rangle$ represents the scalar 
dissipation rate. The linear dependence on $\kappa$ in the expression for $\bar{\epsilon}_{\rm c}$ may 
imply a strong dependence of the scalar dissipation rate on $\kappa$. However, this is not true because 
the scalar gradient also depends on $\kappa$. With decreasing $\kappa$, larger scalar gradients can 
develop, which may compensate the decrease in $\kappa$, as we shall see below. The last term on the 
right hand side of eq.\ (\ref{eq:Cvariance}) is the source term, which corresponds to the increase in scalar 
variance due to the injection of new pollutants. 

To drive turbulence, we used an artificial forcing term ${\bf F}$ in the momentum equation (see 
eqn.~2) which is modeled as a stochastic Ornstein-Uhlenbeck (OU) process \citep{EP88, Benzi+08}
with a user-specified forcing correlation time $t_f$. For all simulations reported here, we drove turbulence 
with solenoidal forcing (i.e., $\nabla\cdot{\bf F}=0$) in the range of wavenumbers, $1 \leq |{\bf k}|L/2\pi \leq 3$ 
such that the average forcing wavenumber is $k_{\rm f}\,L/2\pi \simeq 2$.
The driving scheme can thus be  summarized as 
$\langle F_{i}\left({\bf k}, t\right)\,F_{j}\left({\bf k}, t'\right)\rangle = 
\mathcal{P}_{\rm f}(k)\left(\delta_{ij} - k_{i}k_{j}/k^{2}\right)\,{\rm exp}[-(t-t')/t_{f}]$. 
Our choice of solenoidal forcing is 
motivated by the fact that the dynamo is more efficient for solenoidal driving compared to turbulence driven 
purely by compressive forcing due to more pronounced shearing motions in the former case.
In reality though, turbulence is expected to be of mixed type consisting of both solenoidal and 
compressive motions. 

The random forcing for the scalar field ${\bf T}$ is   identical to the velocity forcing, 
$\langle T\left({\bf k}, t\right)\,T\left({\bf k}, t'\right)\rangle = \mathcal{P}_{\rm T}(k)\,{\rm exp}[-(t-t')/t_{s}]$ 
with $\mathcal{P}_{T}(k)\propto \mathcal{P}_{F}(k)$ and $t_{s}$ is the scalar correlation time. 
In other words, pollutants are continuously injected in our simulations on scales equal to the driving scale 
of the flow. This is similar in spirit to the actual scenario in the ISM where, for mixing of new metals from 
supernovae (SNe), the source length scale is the same as the driving scale of turbulence. On the other 
hand, if we consider self-enrichment in star-forming clouds, the two scales could be different as the driving 
scale of turbulent energy is mainly from the cascade of the interstellar turbulence on scales much larger 
than the scalar source scale which could be within the star-forming cloud \citep[e.g.,][]{Pan+12}. We refer 
the reader to \citet{PS10} for a detailed analysis of the dependence of turbulent mixing depending on 
scalar source size. 

In all our runs, we adopted an isothermal equation of state  and an initial density and sound speed set to unity. 
The initial magnetic field, $B_{0} = (0, 0, B_{0z})$ was in the $z$-direction with a plasma beta, defined as the 
ratio of the thermal and magnetic pressures, varied between $\beta = 2p/B^{2} = 10^{4} - 10^{7}$. These 
values of $\beta$ allowed us to explore the problem at hand within a reasonable computational 
time. Similar to \citet{PS10}, we chose the forcing correlation time for both $t_{f}$ ans $t_{s}$ to be $1/2$ 
the sound crossing timescale, defined as the box size divided by the sound speed. This approach 
ensures that the forcing correlation time is constant for each simulation\footnote{The 
choice of an appropriate forcing correlation may depend on the problem of interest. In some 
cases, numerical studies of small-scale dynamos employ a delta-correlated forcing as in \citet{HBD04}. 
This choice is a simplified choice which may not apply for realistic astrophysical flows. However, since the 
amplification of the magnetic field is driven by eddies below the driving scale, the properties of the small-scale 
dynamo can be expected to have only a minor dependence on the large-scale driving pattern. In fact, the 
small-scale dynamo has also been obtained for a finite time forcing correlation \citep{Cho+09, Fed+11b}. 
In the context of scalar mixing, \citet{PS10} verified that different choices of the forcing correlation did not 
affect the scalar statistics.}. 

Also similar to the study of \citet{PS10}, we used three independent scalars, each of which had 
the same source spectrum, which also matched the spectrum of the velocity forcing. This approach ensures 
smaller temporal variations in the scalar variance compared to the temporal variations of the rms value of the 
concentration of each individual scalar. Moreover, we chose the scalar source term to have a zero mean and 
therefore did not consider the evolution of the mean concentration in our simulations\footnote{Using a Gaussian 
source term is a common practice in the theoretical studies of passive scalars physics, as it isolates the mixing 
physics from the possible complexities arising from the pollutant sources.}.    
The negative concentration values in our simulation are to be understood as relative to the mean concentration. 
In this sense, the negative values of the scalar concentration in our simulations should be interpreted as 
under-enriched regions in the simulation volume.  Thus the addition of large scale regions with negative scalar 
concentration could correspond to an inflow of intergalactic gas onto a galaxy or the collapse of additional 
interstellar gas onto a self-enriched molecular cloud.

All our simulations were performed with a Schmidt number, ${\rm Sc = Pe/Re} = \nu/\kappa = 1$ with constant 
kinematic viscosity and scalar diffusivity. Here ${\rm Pe}$ and ${\rm Re}$ are the Peclet and the fluid Reynolds 
number respectively. In order to explore the parameter space, we adopted a two-fold simulation strategy.  First, 
we explored turbulent mixing as a function of the rms Mach number ($\mathcal{M}_{\rm rms} \approx 0.3, 1.1, 2.3$) 
for a constant Prandtl number of unity (i.e., ${\rm Pm = Rm/Re} = \nu/\eta = 1$) at ${\rm Re = Rm} \approx 1250$. 
Here ${\rm Rm} = u\,l_{\rm f}/\eta$ is the magnetic Reynolds number with $u$ being the rms velocity and 
$l_{\rm f} = 2\pi/k_{\rm f}$ the driving scale of turbulence. Next, we explored the mixing of pollutants as a 
function of the Prandtl number for a fixed Mach number. To this end, we conducted simulations at 
${\rm Pm} = 1, 2$ and $4$ for a subsonic rms ${\mathcal M} = 0.3$. To achieve ${\rm Pm}>1$, we increased the 
kinematic viscosity $\nu$ (thereby decreasing ${\rm Re}$) instead of decreasing the magnetic diffusivity $\eta$ 
so as to ensure that most of the scales of the velocity and the magnetic field remained resolved at $512^{3}$ 
resolution. Furthermore, this approach helps us to probe if there are qualitative differences in the turbulent 
mixing depending on whether the flow field is turbulent or random. In addition, we also conducted two ideal 
MHD simulations at $\mathcal{M}=0.3$ with initial plasma beta $\beta_{\rm in}=10^{4}$ and $10^{7}$ respectively 
and another ideal run at $\mathcal{M}=2.4$ with $\beta_{\rm in}=10^{7}$. In these runs, the viscous and the diffusive 
scales are determined by the numerical scheme rather than the user,  although larger inertial ranges are achieved. 
In all cases, the magnetic Reynolds number ${\rm Rm} > {\rm R}_{\rm cr}$ ensuring magnetic field growth by 
dynamo action. Table~\ref{sumsim} provides a summary of the simulation runs we performed. 

%%%%%%%%%%%%%%%%%%%%%%%%%%%%%%%%%%%%%%%%%%%%%%%%%%%%
\begin{table}[h]
\begin{minipage}{92mm}
\begin{center}
\resizebox{0.98\columnwidth}{!}{%
\begin{tabular}{|c|c|c|c|c|} \hline \hline 
Simulation & Init. Plasma  & Rms Mach & Prandtl \# & Reynolds \# \\
Run & Beta, $\beta_{\rm in}$ & $\mathcal{M}_{\rm rms}$ & ${\rm Pm = Rm/Re}$ & ${\rm Re} = u\;l_{f}/\nu$ 
\\ \hline \hline
M0.3Pm1a & $10^{4}$ & $\approx 0.3$ & 1 & $\approx 1250$ \\ \hline
M0.3Pm1b & $10^{7}$ & $\approx 0.3$ & 1 & $\approx 1250$ \\ \hline
M0.3Pm2 & $10^{4}$ & $\approx 0.3$ & 2 & $\approx 620$ \\ \hline
M0.3Pm4 & $10^{4}$ & $\approx 0.3$ & 4 & $\approx 312$ \\ \hline
M1.1Pm1 & $10^{4}$ & $\approx 1.1$ & 1 & $\approx 1250$ \\ \hline
M2.3Pm1 & $10^{4}$ & $\approx 2.3$ & 1 & $\approx 1400$ \\ \hline
M0.3Id1 & $10^{4}$ & $\approx 0.3$ & $\approx O(1)$ & ------ \\ \hline 
M0.3Id2 & $10^{7}$ & $\approx 0.3$ & $\approx O(1)$ & ------ \\ \hline 
M2.4Id   & $10^{7}$ & $\approx 2.4$ & $\approx O(1)$ & ------ \\ \hline \hline
\end{tabular}
}
\end{center}
\caption{Summary of the simulation runs presented in this paper at a uniform grid resolution of 
$512^{3}$. The non-ideal MHD runs are classified as M{\bf X}Pm{\bf Y} where {\bf X} refers to the 
value of the Mach number and {\bf Y} refers to the value of Prandtl number. M2.3Pm1 is only 
followed up to the linear growth phase. M0.3Id1 and M0.3Id2 are the two ideal MHD runs at 
$\mathcal{M}=0.3$ with $\beta_{\rm in} = 10^{4}$ and $10^{7}$ respectively. M2.4Id corresponds 
to an ideal run at $\mathcal{M}=2.4$ with $\beta_{\rm in}=10^{7}$. In all the runs, the Schmidt 
number ${\rm Sc}=\nu/\kappa=1$.}
\label{sumsim}
\end{minipage}
\end{table}
%%%%%%%%%%%%%%%%%%%%%%%%%%%%%%%%%%%%%%%%%%%%%%%%%%%%%%%%%

\section{Results}\label{results}

\subsection{Time Evolution}

The general properties of the small-scale dynamo have been extensively studied with the help of direct 
numerical simulations \citep[][and other references therein]{Scheko+02, Scheko+04,HBD04,BS05, Cho+09, 
CR09, Fed+11b, BSS12, Beres12, BS13} for both incompressible and compressible turbulence. The magnetic 
field is amplified by random stretching and folding by the turbulent eddies. This occurs exponentially at first, in 
what is labeled the kinematic phase, with the magnetic energy peaking on scales nearer to or somewhat 
larger than the dissipation scale. Once energy equipartition is attained on these scales, the stretching of 
the magnetic field lines is hindered by the magnetic back reaction and an intermediate stage of linear growth 
follows. At this stage, the peak of the magnetic power spectrum begins to shift as the magnetic field attains 
equipartition on increasingly larger scales. Eventually, when the magnetic energy density becomes 
comparable to the kinetic energy density close to the driving scale of turbulence, field amplification by the 
small-scale dynamo becomes impossible, and the system enters the saturated phase.

Our main objective here is to explore how the mixing of pollutants evolve as the magnetic field builds up 
due to small-scale dynamo action. In the kinematic phase we expect the mixing to be similar to that of a 
hydrodynamical simulation as the magnetic field is dynamically unimportant. To explore this, we conduct 
an ideal and a non-ideal run at $\mathcal{M}=0.3$ and an ideal run at $\mathcal{M}=2.4$ with 
$\beta_{\rm in}=10^{7},$ which allow for a kinematic growth phase sufficient for our purpose. However, to 
break new ground, it is more important to investigate the effect of the magnetic field on the scalar mixing in 
the regime in which the magnetic field becomes dynamically important to back react on the flow, as will 
occur in any continuously enriched and driven magnetized turbulent medium. Therefore, we conduct most 
of our simulations with $\beta_{\rm in}=10^{4},$ sacrificing the majority of the kinematic regime in favor of 
probing the mixing properties in the saturated phase within reasonable computational time and resources. 

\begin{figure}[h]
%begin{center}
\hspace{-0.22in}
\includegraphics[width=1.08\columnwidth]{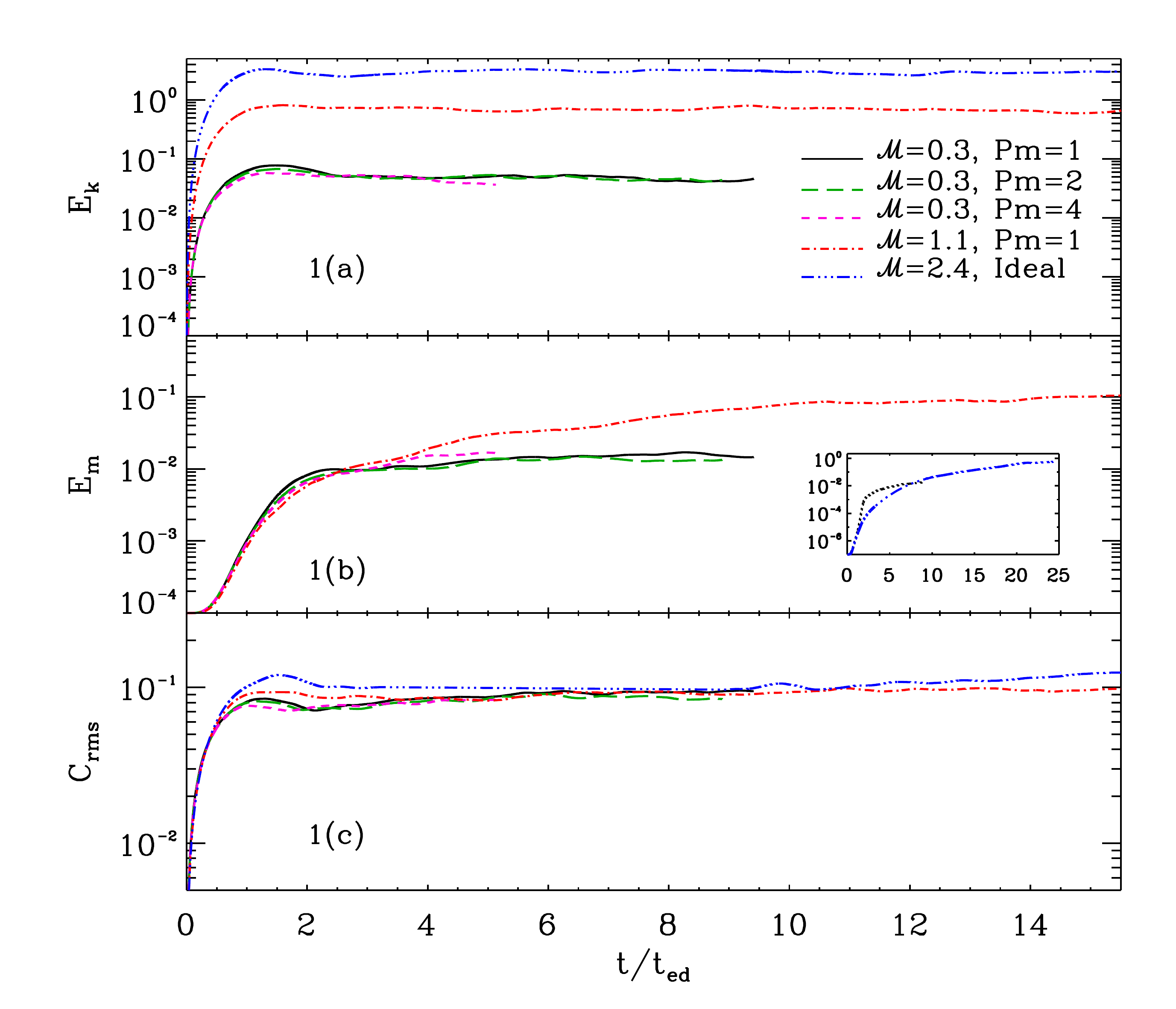}
\vspace{-0.2in}
%\end{center}
\caption{Time evolution of the kinetic energy (panel 1a), magnetic energy (panel 1b), and the rms value 
of the scalar concentration (panel 1c) for the different MHD runs starting with $\beta_{\rm in}=10^{4}$ 
and $10^{7}$. The inset figure in panel 1(b) shows the growth of the magnetic energy for M0.3Id2 
(black, dotted) and M2.4Id (blue, dash-dotted) runs. 
\label{tseries1}}
\end{figure}

Figure~\ref{tseries1} shows the evolution of the kinetic and magnetic energies and the rms value of the 
scalar concentration field as a function of the eddy turnover time $t_{\rm ed}=l_{\rm f}/u_{\rm rms}$.  
The presence of a very short kinematic phase does not allow us to make statements about the growth rate 
of the magnetic field. Nevertheless, in the range $t=(1-2.5)\,t_{\rm ed}$, panel 1(b) shows the decrease in 
growth rate with increasing Mach number. The inset figure in panel 1(b) shows more clearly the diminishing 
growth rate of the magnetic field as the Mach number increases from $\mathcal{M}=0.3$ to $2.4$ for the 
ideal runs, which start from an initial $\beta=10^{7}$. For solenoidal forcing, such a decrease in growth rate 
as the turbulent flow becomes supersonic has been previously reported in direct simulations of the 
small-scale dynamo \citep{Fed+11b}. However, they also claimed the growth rate starts to increase for 
$\mathcal{M}>2.4$, which we do not address here. 

In our simulations, the growth rate also seems to decrease with increasing ${\rm Pm}$ (and decreasing 
${\rm Re}$ and ${\rm Pe}$), although a more extended kinematic range is required to settle this issue. 
The transition to the intermediate stage of magnetic field growth at the end of the kinematic phase is clearly 
seen for $\mathcal{M}=1.1$ where it extends from $t\approx(4-10)\,t_{\rm ed}$, while for $\mathcal{M}=2.4$, the 
intermediate growth phase is $t\approx(10-20)\,t_{\rm ed}$. The magnetic field finally saturates at a level 
depending on the Mach number of the flow. We note here that the high Mach number 
runs are typically more expensive than their subsonic counterparts. This is mainly due to two reasons. First, 
the resulting time step becomes smaller from subsonic to supersonic due to the Courant-Friedrichs-Lewy (CFL) 
condition and secondly, the presence of supersonic motions means that a substantial fraction of the kinetic 
energy is contained in shocks, which are relatively inefficient at driving the dynamo. This leads to a reduced 
growth rate of the magnetic field. In panel 1(c), we plot the time evolution of the rms value of the scalar concentration 
field for the above runs. The evolution of the rms value of the scalar concentration shows that $C_{\rm rms}$ 
increases as the flow moves from the kinematic phase to the saturated phase.    Furthermore,
in both the kinematic and saturated phases, $C_{\rm rms},$  increases with increasing Mach number,  
but  is largely independent of ${\rm Pm}$.
 
%%%%%%%%%%%%%%%%%%%% Time evolution %%%%%%%%%%%%%%%%%%%%%%%
%%%%%%%%%%%%%%%%%%%%%%%%%%%%%%%%%%%%%%%%%%%%%%%%%%%%%%%%

%%%%%%%%% Figure 1 : three dimensional snapshots %%%%%%%%%%%%%%%%%%%%%%%%%%%
\begin{figure*}
\begin{center}
\includegraphics[width=2.1\columnwidth]{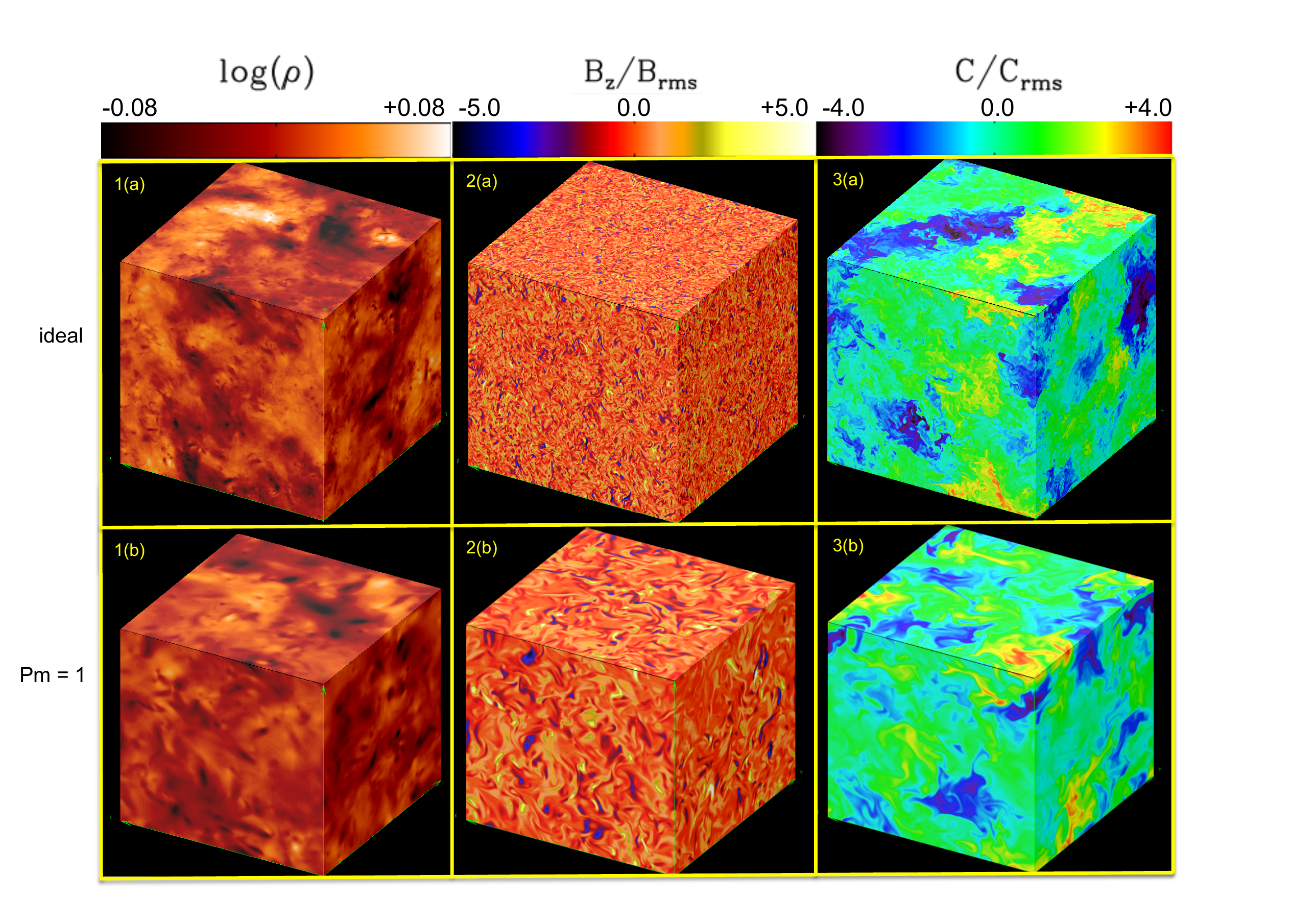}
\end{center}
\vspace{-0.1in}
\caption{Three dimensional volume rendering from the ideal (first row) and the non-ideal (second row) 
runs at $\mathcal{M}=0.3$, $\beta_{\rm in}=10^{7}$, M0.3Id2 and M0.3Pm1b.  The panels show
 showing the density (panels 1a and 1b), the $z$-component 
of the magnetic field normalized by the rms magnetic field ${\rm B_z/B_{rms}}$ (panels 2a and 2b) and 
the scalar concentration field normalized by the rms value (panels 3a and 3b) on the periphery of the 
simulation volume at $t\simeq2.5t_{\rm ed}$. This corresponds to the kinematic phase of the dynamo. 
Note the initial weak vertical magnetic field is deformed by the underlying turbulence and resembles 
white noise. 
\label{3dkin}}
\vspace{0.05in}
\end{figure*}
%%%%%%%%%% Figure 2 : three dimensional snapshots in the saturated state %%%%%%%%%%%%%
\begin{figure*}
\begin{center}
\includegraphics[width=2.1\columnwidth]{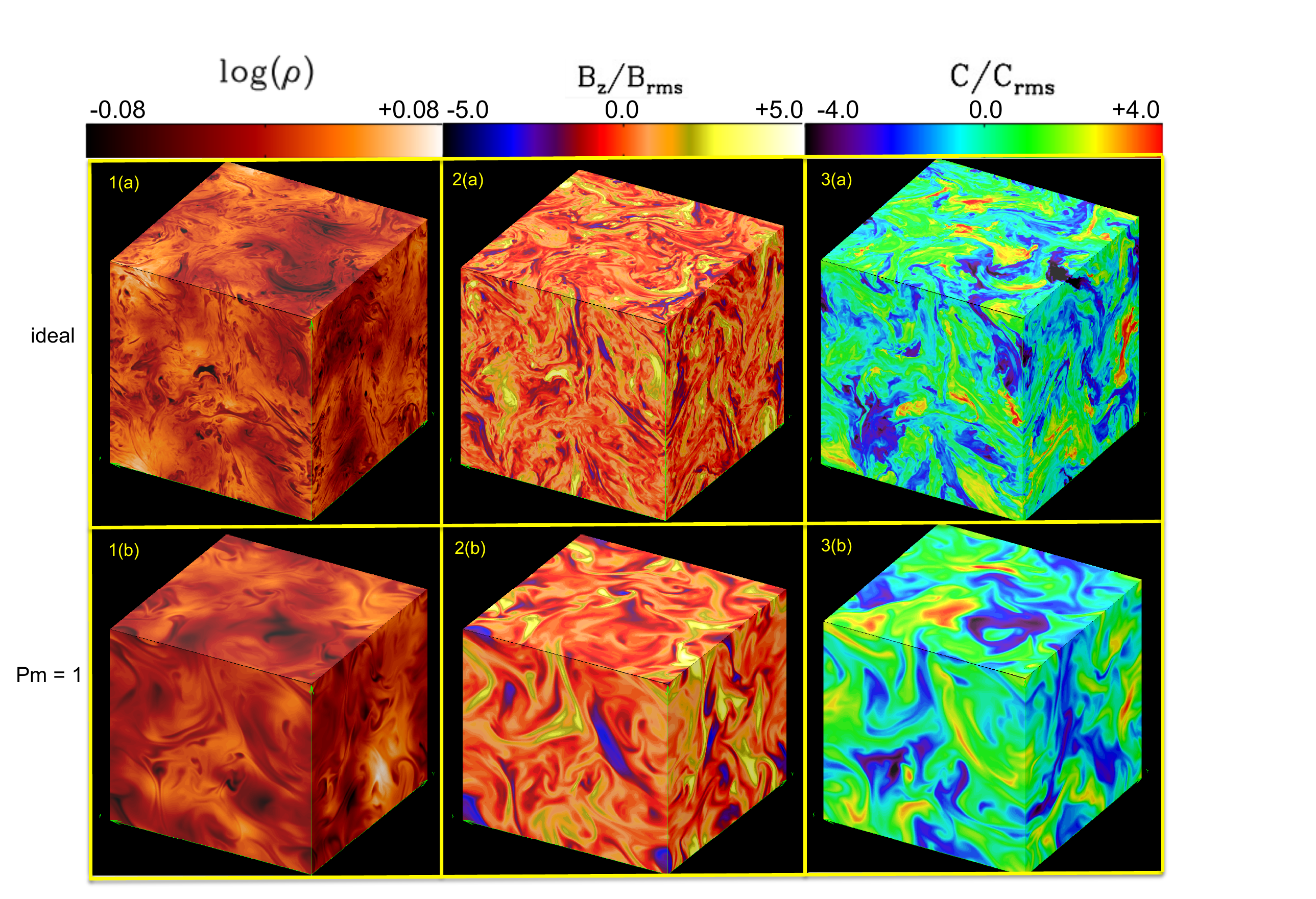}
\end{center}
\vspace{-0.1in}
\caption{Same as in Fig.~\ref{3dkin} but now at $t\simeq8.8t_{\rm ed}$ when the 
magnetic field is saturated, obtained from runs M0.3Id1 and M0.3Pm1a.
Note the presence of long filamentary structures in both the 
magnetic field and the scalar concentration field compared to the ones in Fig.~\ref{3dkin}. 
\label{3dsat}}
\vspace{0.05in}
\end{figure*}
%%%%%%%%%%%%%%%%%%%%%%%%%%%%%%%%%%%%%%%%%%%%%%%%%%%%%%

\subsection{Overall Field Structure}

In Fig.~\ref{3dkin} we show three-dimensional volume renderings of the density, magnetic field, and the 
scalar concentration field for both the ideal and the non-ideal $\mathcal{M}=0.3$ runs at  $t\approx2.5t_{\rm ed},$ 
corresponding to the kinematic phase when the magnetic field is too weak to affect the flow. Fig.~\ref{3dsat}, 
shows similar renderings of the same fields at $t\approx8.8t_{\rm ed}$ corresponding to the saturated state 
when nonlinear back reactions resist  further stretching and folding of the field by the turbulent eddies. 

Several qualitative differences are apparent in these plots. In the non-ideal MHD run, magnetic and scalar 
structures  are thick compared to the ideal MHD run, due to the inclusion of explicit resistivity and diffusion. 
As explained in the previous section, we start with a weak vertical magnetic field that then gets randomly 
stretched and folded as the turbulence is stirred in the box. In the kinetic phase, the magnetic field is like 
white noise, and from panels 2a and 2b in Fig.~\ref{3dkin}, we see that it appears to have 
strong positive/negative values (denoted by yellow/blue) in only a few regions in the simulation volume 
where random stretching by turbulent eddies is efficient in growing the field. At late times, on the other 
hand, the coherence length of the magnetic field grows to near the driving scale, saturating in a configuration 
shown in panels 2a and b in Fig.~\ref{3dsat}. 

Similar to the magnetic field, the scalar field also shows numerous small-scale structures with sharp 
concentration contrasts  in the kinematic phase resulting  from random stretching and shearing by turbulent 
eddies (clearly visible in panel 3a in Fig.~\ref{3dkin}). Such sharp contrasts, unlike shocks, are not 
discontinuous, but rather have a small, but significant thickness.  They are usually referred to as `cliffs and ramps' 
in incompressible passive scalar turbulence \citep{Sreeni96,SS00}. They are produced by turbulent stretching, 
which amplifies scalar gradients by bringing fluid elements with very different concentration levels next to each 
other. As the magnetic field saturates, the morphology of the scalar structures changes to long ribbon-like filaments 
as shown in panels 3a and 3b in Fig.~\ref{3dsat}. This appears to be an effect of the magnetic field and was not 
seen previously in the hydrodynamical simulations of \citet{PS10}. 
A quantitative analysis of the properties of the scalar structures is presented in the 
next subsection.

\subsection{Power Spectra and Structure Functions}

\subsubsection{Power Spectra}

%%%%%%%%%%%%%%%%%%% Fig.3, power spectra for the different runs %%%%%%%%%%%%%%%%%
\begin{figure*}
\begin{center}
\includegraphics[width=2.05\columnwidth]{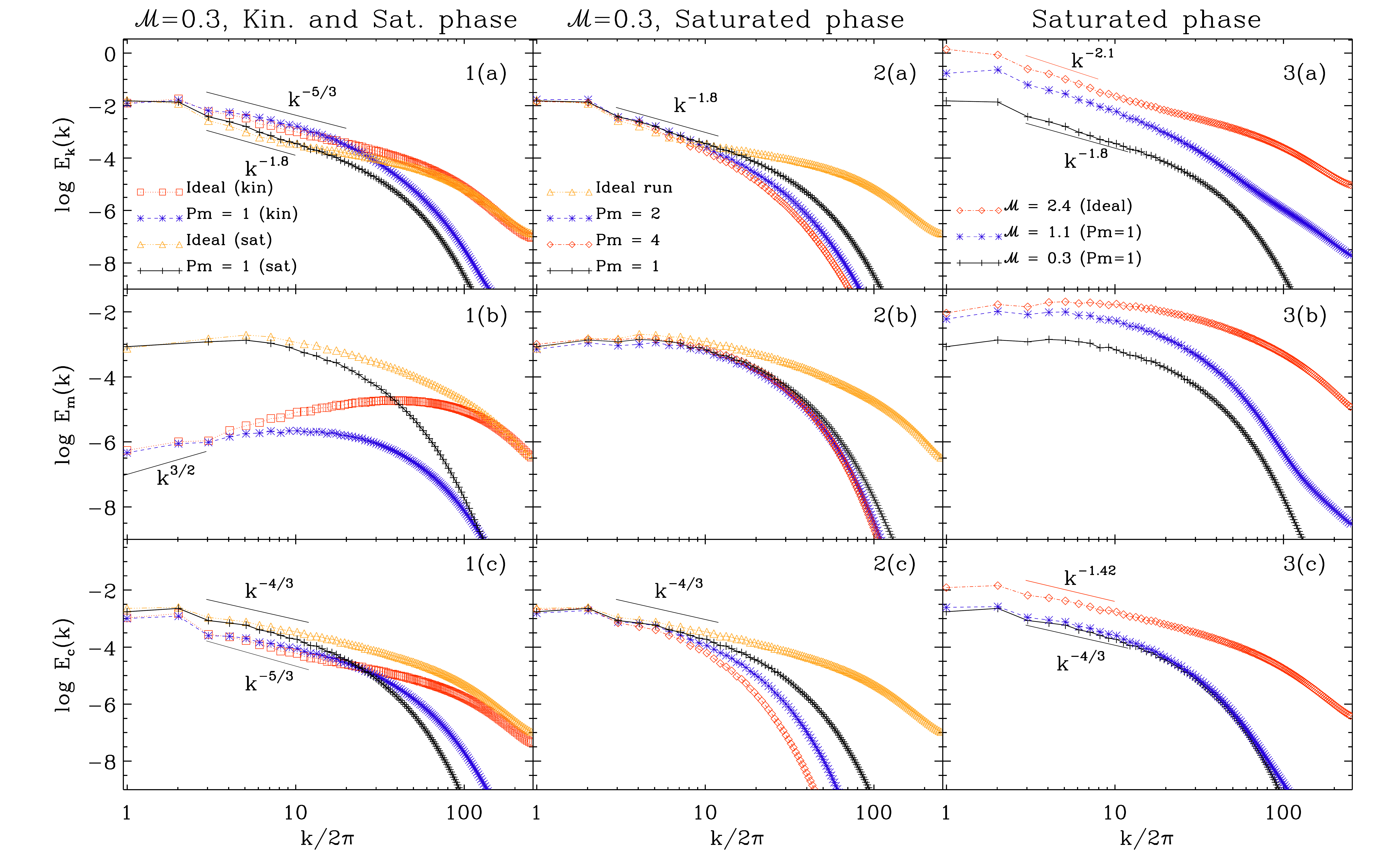}
\end{center}
\vspace{-0.2in}
\caption{Power spectra of the velocity, magnetic and the scalar field - for the kinematic and saturated 
phase of evolution at $\mathcal{M}=0.3$ (column 1), as a function of different Prandtl numbers for $\mathcal{M}=0.3$ 
in the saturated state of the dynamo (column 2) and as function of different Mach numbers for a Prandtl number 
of unity (column 3). The errors in the estimate of the power spectra are less than $10\%$ of their values. 
\label{spectra}}
\vspace{0.1in}
\end{figure*}
%%%%%%%%%%%%%%%%%%%%%%%%%%%%%%%%%%%%%%%%%%%%%%%%%%%%%%

In Figure~\ref{spectra}, we plot the three-dimensional power spectra of the velocity, magnetic and the scalar 
concentration field for the different  runs, in both the kinematic and saturated phases of evolution. 
In the first column (panels 1a, 1b and 1c), we compare the power spectra obtained from the ideal MHD run 
to that of the non-ideal run at $\mathcal{M}=0.3$. Because the viscous scale is larger for the non-ideal run, 
the turnover point for this run lies at smaller $k.$  Nevertheless, over the 
inertial range, the velocity power spectra for both the ideal and the non-ideal run have a $\approx k^{-5/3}$ 
scaling in the kinematic phase (dotted red and blue dashed lines), measured from runs with $\beta_{\rm in}=10^{7}.$
The power spectrum peaks at $k/2\pi=2$ which corresponds to the energy injection scale in our simulations. 
In the saturated phase, the velocity spectrum has a slightly steeper scaling in the range $k\sim(3-9)$ 
which can be attributed to energy transfer from $E_{\rm k}$ to $E_{\rm m}$ which occurs faster at smaller 
scales. A simple linear regression analysis gives the best-fit slope to be $\approx k^{-1.85}$. 

The magnetic power spectra plotted in panel 1(b) show the evolution in the kinematic and saturated phases 
of the dynamo. At low $k$, the magnetic field spectra are flatter than the Kazantsev $k^{3/2}$ scaling at early 
times \citep{K68, BS05, BS13} although still peaked at much larger scales at $k\approx20$ for the non-ideal run 
and at about $k\approx40$ for the ideal run. This difference is again because of the larger inertial range available 
in the ideal case which leads to a lot more eddies amplifying the field at smaller scales compared to the non-ideal 
run at ${\rm Pm}=1$. If $u_{l}$ is the velocity fluctuations on a scale $l$, then $t_{\rm ed} = l/u_{l} \propto l^{2/3}$ 
for Kolmogorov turbulence. Thus, smaller eddies amplify the magnetic field faster because of their shorter turnover 
times and hence the magnetic power spectrum initially peaks at large $k$. However, as the magnetic field 
approaches saturation, the peak of the magnetic spectrum shifts to larger scales at about $k\approx4$, implying 
an increase in the coherence length of the field. This feature can also be seen in the three-dimensional volume 
rendering plots upon comparing panels 2(b) in Figs.~\ref{3dkin} and \ref{3dsat}. 

In Figure 4, panel 1(c) shows the time evolution of the scalar power spectrum. The classical picture for the 
generation of scalar structures at small scales is a cascade process that is similar to the cascade of the velocity 
field, resulting in a $k^{-5/3}$ scalar spectrum for isotropic incompressible turbulence \citep{Sreeni96, WG04}. 
Since we use three independent scalars, the power spectrum is averaged over the three values at each 
$k,$ increasing the accuracy of our measurements. For the modest Reynolds numbers $\approx 1250$ achieved 
in our simulations, we find a $k^{-5/3}$ scaling in a narrow band of wave numbers $k\approx(3-8)$ in the kinematic 
phase which then becomes flatter $\approx k^{-4/3}$ in the saturated phase. We note here that measuring the 
scalar spectrum slope accurately is somewhat ambiguous because of the sudden drop in the spectrum from 
$k=2$ to $3$. The best fit scaling seems to be more like $k^{-1.15}$ for the saturated phase. 

Comparing panels 1(a) and 1(c) we find that the slope of the velocity spectra decreases as the system evolves 
from the kinematic to the saturated state. However, the slope of the scalar power spectrum increases. As the 
system approaches saturation, the Lorentz forces become strong enough to affect the velocity field thereby reducing 
the velocity fluctuations. This leads to an increase in scalar fluctuations and hence a flatter scalar power spectrum. 
If $\alpha_{u}$ and $\alpha_{c}$ are the scaling exponents of the velocity and the scalar power spectrum respectively, 
then the OC theory predicts
\EQ
\alpha_{c} = -\left[\frac{5 + \alpha_{u}}{2}\right],
\label{alpc}
\EN
as described in more detail below.
In the kinematic phase, for $\alpha_{u} = -5/3$, the above relation predicts $\alpha_{c} = -5/3$ which is in 
perfect agreement with the slopes obtained in our simulations. In the saturated phase where $\alpha_{u} = -1.85$ 
as obtained from our best-fit model, $\alpha_{c}= -1.57$ which is slightly steeper than the 
$\approx k^{-1.15}$ scaling we obtain in our simulations.  Note however that because the inertial range is small in 
our simulations and, more importantly, because $E_k(k)$ and $E_c(k)$ in the saturated phase may not be strictly 
power laws, a perfect match between these models is not expected.   Similarly, the best-fit $k$ space 
power-law and the structure function power law fits below are expected to show the same trends, but not 
necessary exactly obey $\alpha = -(1 + \xi)$, where $\xi$ is the structure function slope, as pure 
power laws would do.

In the second column, we plot the velocity, magnetic and the scalar power spectra for runs with 
$\mathcal{M}=0.3$ in the saturated state for different Prandtl numbers. As stated earlier, for our high ${\rm Pm}$ 
runs, we decrease the fluid Reynolds number, keeping the magnetic Reynolds number unchanged. Therefore, 
as we go from ${\rm Pm}=1$ to $4$, the viscous scale moves to smaller $k$ values for both the 
velocity and the scalar spectra as evident from panels 2(a) and 2(c). However, the slope of the velocity
power spectrum appears to be independent of ${\rm Pm}$ and has the same $\approx k^{-1.8}$ scaling in the 
range $k\approx 3-8$ for the range of magnetic Prandtl numbers explored in this study. One would expect the 
velocity spectra to become steeper with increasing ${\rm Pm}$ as the fluid becomes increasingly viscous. 
On the other hand, the dissipation scale for the magnetic field remains independent of the change 
in ${\rm Pm}$ because {\rm Rm} is the same for all the runs. In the case of the scalar 
concentration field, the slope of the spectrum is flatter $\approx k^{-4/3}$ for the ideal and the ${\rm Pm}=1, 2$ 
non-ideal runs in the range $k\approx 3-7$ while for ${\rm Pm} = 4$, the above scaling holds only for a very 
short range of $k\sim3-5$. 

In the third column, we show the power spectra in the saturated phase for the different Mach numbers. 
In particular, we plot the velocity, magnetic and the scalar spectra for M0.3Id2, M2.4Id and M0.3Pm1a runs. 
As evident from panel 3(a), the velocity power spectrum for $\mathcal{M}=2.4$ shows a $\approx k^{-2.1}$ 
scaling in the saturated phase. In the kinematic phase, the velocity spectrum shows a $\approx k^{-2}$ scaling 
resulting from the dominance of shocks in the turbulent flow \citep[e.g.][]{PS10}. Together with the increase in 
shock intensity with increasing Mach numbers and the fact that the pressure term preferentially converts kinetic 
to thermal energy thereby resulting in the loss of kinetic energy along the cascade causes the velocity spectrum 
to be much steeper than the subsonic case. However, the steepening of the spectrum in the saturated phase 
has to do with the velocity fluctuations being dominated by the magnetic field leading to the familiar behavior 
seen in our saturated runs (see panel 2a). In this phase, the scalar power spectra shows a $k^{-1.42}$ 
scaling for $\mathcal{M}=2.4$ while a $k^{-1.4}$ for the other two runs. The best-fit scaling for 
$\mathcal{M}=2.4$ is $\approx k^{-1.44}$ while the scaling predicted from eq.~(\ref{alpc}) is $\approx k^{-1.45}$. 

In addition, we also computed the evolution of the integral scale of the velocity, magnetic and the scalar 
concentration field for the simulation runs presented here. The velocity and the magnetic integral scale evolution 
has been previously reported in the works of \citet{CR09} and \citet{BS13}. We find that while the velocity and the 
scalar integral scale remains almost constant, the integral scale for the magnetic field increases by a factor of 
$\sim 3-4$ as the system evolves from the kinematic to the saturated phase. This increase is consistent with the 
results reported in \citet{CR09} and \citet{BS13} and can explain the gradual increase in coherence scale of the 
magnetic field structures seen by comparing Figs.~\ref{3dkin} and \ref{3dsat}.

 \subsubsection{Structure Functions and Dimensionality of Scalar Structures}

\begin{figure}[h]
%\begin{center}
\hspace{-.2in}
\includegraphics[width=1.09\columnwidth]{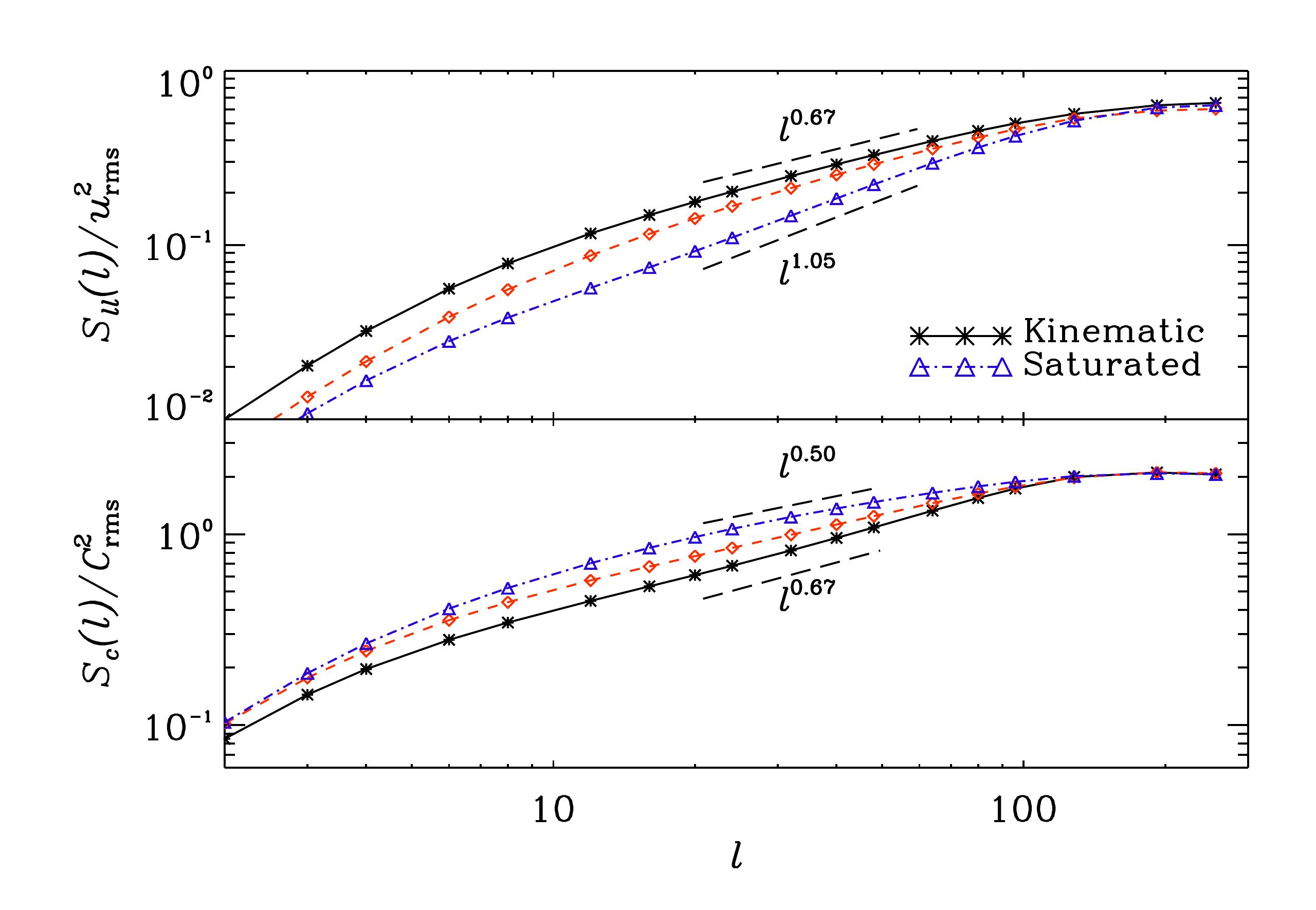}
%\end{center}
\vspace{-.2in}
\caption{Second order longitudinal velocity structure function (upper panel) and the scalar structure 
function (lower panel) for run M0.3Id2 in the kinematic (black solid), intermediate (red dashed) and 
saturated (blue dash dotted) phases. The velocity field scales as $l^{2/3}$ in the kinematic phase 
and as $l^{1.05}$ in the saturated phase. The scalar concentration field also shows an $l^{2/3}$ 
scaling in the kinematic but an $l^{1/2}$ in the saturated phase consistent with the OC picture. 
\label{stf}}
\end{figure}

In this subsection, we analyze the scaling properties of the second order longitudinal velocity structure 
functions (SFs), defined as $S^{u}_{||}(l)\equiv \langle[(\uu(\bfx + \bfl) - \uu(\bfx))\cdot \bfl/l]^2\rangle$ and 
the scalar structure function $S_{C}(l)\equiv \langle(C(\bfx + \bfl) - C(\bfx))^2\rangle$, where 
$l$ is the separation distance.
Figure~\ref{stf} shows a plot of  $S^{u}_{||}(l)$ (upper panel) and  $S_{C}(l)$ (lower panel) in our M0.3Id2 run,  
for three different phases. In the kinematic phase, when the magnetic field is dynamically 
unimportant, the velocity structure function shows an $l^{2/3}$ scaling, in perfect agreement with the predictions 
of the Kolmogorov theory. In the saturated phase, this scaling steepens to $\approx l^{1.05}$. A similar behavior 
was also observed from the velocity power spectra plots in Fig.~\ref{spectra} which shows the slope of the 
velocity spectra to be slightly steeper than $k^{-5/3}$ in the saturated regime. Physically this is due to the transfer 
of the kinetic to magnetic energy,  and it occurs faster at smaller scales, leading to a steepening of 
the velocity structure function or spectra. A supporting evidence is that the structure function, 
$\langle[({\bf u}(\bfx + \bfl) - {\bf u}(\bfx))\cdot \bfl/l]^2\rangle +\frac{1}{4\pi} \langle[({\bf B} (\bfx + \bfl) - {\bf B}(\bfx))\cdot \bfl/l]^2\rangle$, 
corresponding to the sum of the kinetic and magnetic energies, is much shallower, $\propto l^{0.60}$, 
than the velocity structure function itself. The scaling exponent of $S^{u}_{||}(l)$ in the intermediate phase 
(red dashed curve)  lies somewhere in between the above two values for the kinematic and saturated phases.

%%%%%%%%%%%%%%%%%%%%%%%%%%%%%%%%%%%%%%%%%%%%%%%%%%%%%%
%%%%%%%%%%%%%%%%%%%%%%%%%%%%%%%%%%%%%%%%%%%%%%%%%%%%%%

The scalar evolution is regulated by the velocity field  (see eqn.~\ref{eq:Cvariance}), and, as a response to 
the changes in the velocity statistics in two phases, the scalar structure function, $S_C (l)$, also shows different 
scalings, flattening  from $l^{2/3}$ in the kinematic phase to $l^{0.5}$ in the saturated phase. We find that 
these scaling exponents are consistent with the OC picture, where a constant flux of scalar energy occurs 
along the cascade over the inertial range. In this range, the scalar fluctuations are determined by the advecting 
velocity only. If we assume that the timescale for a cascade step is of $\approx O(t_{\rm ed})$ at a given scale, 
then the difference in scalar concentration over a scale '$l$' is related to the velocity difference as 
\EQ
\delta C(l)^{2} \simeq {\overline{\epsilon}}_{c}\frac{l}{\delta u(l)}, 
\label{cvrel}
\EN
where $\overline{\epsilon}_{c}$ is the scalar dissipation rate. Therefore, if the second order velocity and 
scalar structure functions are defined as, 
$\langle \delta u(l)^{2}\rangle \propto l^{\xi_u},$ and
$\langle \delta C(l)^{2}\rangle \propto l^{\xi_c},$ 
the scaling exponents are related as 
\EQ
\xi_{c} \simeq 1 - {\xi_{u}}/{2}. 
\label{xic}
\EN
Substituting $\xi_{u} = (0.67, 1.05)$ obtained from Fig.~\ref{stf}, the above relation predicts 
$\xi_{c} = (0.665, 0.475),$  in good agreement with $\xi_{c} = (0.67, 0.50)$ inferred 
from our simulations as shown in the figure. Repeating this analysis for our supersonic 
ideal MHD run, M2.4Id we find $\xi_{u} = (0.85, 0.92)$ and $\xi_{c} = (0.56, 0.52)$ which 
are again in good agreement with the scaling exponents obtained from eq.~\ref{xic}.
However, since the inertial range obtained from our structure functions are quite narrow even 
for our ideal MHD runs, we refrain from computing the same for our non-ideal MHD runs which 
has an even narrower inertial range due to the modest Reynolds numbers achievable in our 
simulations. 

We now turn our attention to the nature of the hierarchical structures of the velocity and 
scalar fields by analyzing structure functions at high orders.  At the outset, we define the velocity, 
scalar and the mixed SF's of order $p$ as, 
\EQA
S^{u}_{p}(l) &\equiv& \langle|\delta u(l)|^p\rangle \propto l^{{\xi_u}(p)},
S^{C}_{p}(l) \equiv \langle|\delta C(l)|^p\rangle \propto l^{{\xi_C}(p)}, \nonumber \\
S^{m}_{p}(l) &\equiv& \langle|\delta u(l)\,\delta C(l)^2|^{p/3}\rangle \propto l^{\xi_{m}(p)}, 
\ENA
where $\delta u(l)$ and $\delta C$ are the longitudinal velocity and scalar increments at a distance of $l$.
The She-Leveque (SL) \citep{SL94} intermittency model gives a prediction for the scaling exponent $\xi(p)$, 
\EQ
\xi(p) = \gamma p + C(1-\beta^p_{i}), 
\EN 
where $\beta_{i}$ is the intermittency parameter, $\gamma$ is the scaling exponent for the most intense 
structures, and $C$ is interpreted as the codimension of these structures, such that fraction dimension is
$d=3-C$ for three-dimensional flows. 
The model can thus be used to study the geometry of the strong velocity or scalar structures.  
Because of the limited inertial range available in our turbulence simulations, it is difficult to 
measure $\xi(p)$ at high orders. In such cases, it turns out that one can exploit the self-similarity 
hypothesis of \citet{Benzi+93} to measure the SF's at all orders against the third order ones, which allows for 
an extended power-law range and more accurate measurements. We therefore measure the scaling exponents 
$\zeta(p)$ relative to the third order SF's defined as $S_{p}(l) \propto [S_{3}(l)]^{\zeta(p)}$. 
By definition, $\zeta(p) = \xi(p)/\xi(3)$ and $\zeta(3) = 1$. 
%%%%%%%%%%%%%%%%%%%%%%%%%%%%%%%%%%%%%%%%%%%%%%%%%%%%%%
\begin{figure}[h]
%\begin{center}
\hspace{-.28in}
\includegraphics[width=1.13\columnwidth]{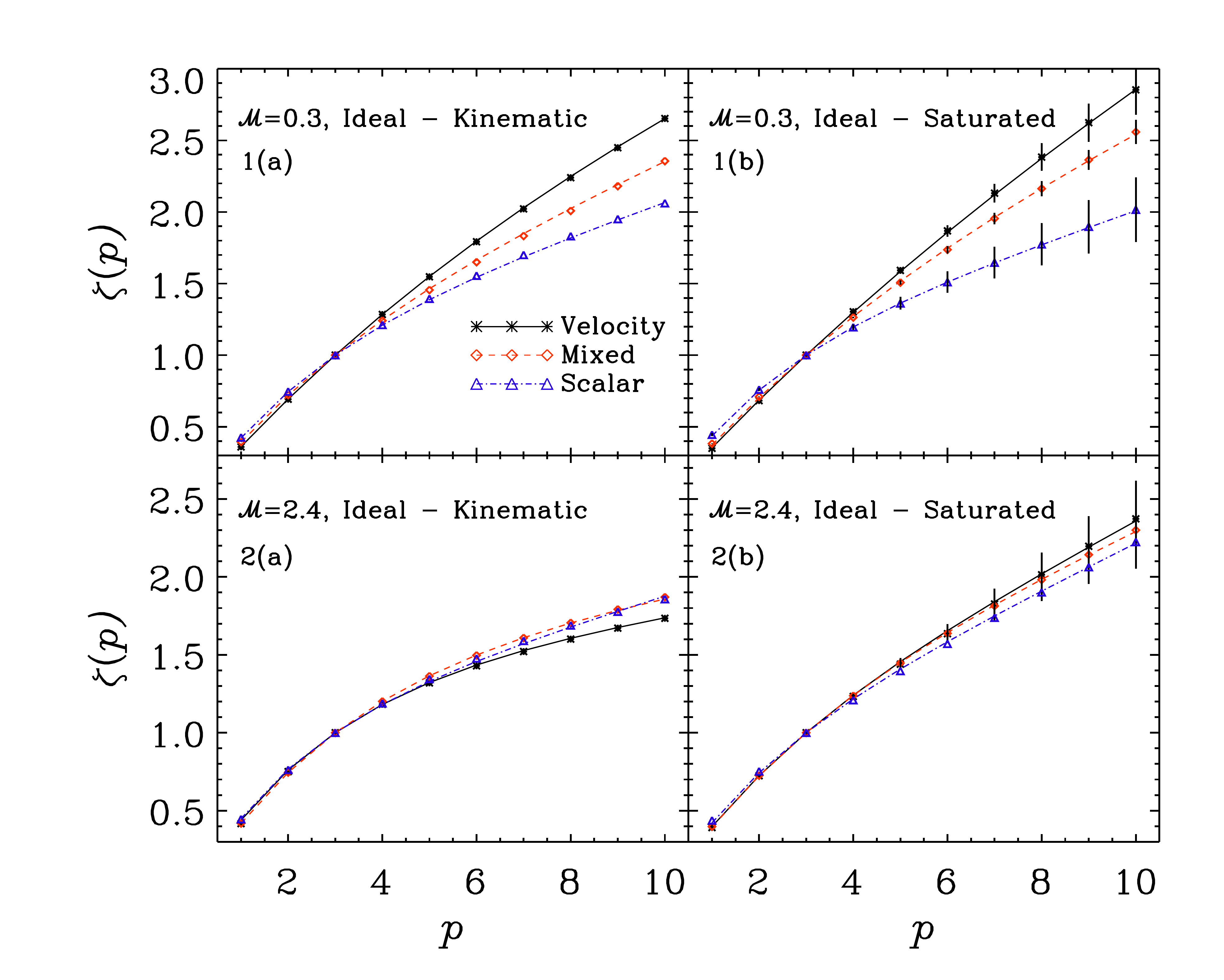}
%\end{center}
\vspace{-.2in}
\caption{Scaling exponents $\zeta_{p}$ for the SF's in the kinematic and saturated phases for runs : M0.3Id2 
(upper row) and M2.4Id (bottom row). 
Black vertical lines in the panels depict the error bars. The scalar field appears to be more intermittent 
than the velocity in both the phases for $\mathcal{M}=0.3$ while the velocity appears to be more intermittent 
at $\mathcal{M}=2.4$. 
\label{dim}}
\end{figure}
%%%%%%%%%%%%%%%%%%%%%%%%%%%%%%%%%%%%%%%%%%%%%%%%%%%%%%

In Fig.~\ref{dim}, we plot the scaling exponents $\zeta_{p}$ for the velocity, scalar and the mixed SF's 
for the ideal runs with $\mathcal{M}=0.3$ (upper row) and $2.4$ (bottom row). 
We find that for all $p$, the values of $\zeta_{p}$ are consistent with the generalized SL model, 
\EQ
\zeta(p) = \gamma'\,p + C'\,(1-\beta^p_{i}), 
\label{sleqn}
\EN
where $\gamma' = \gamma/\xi(3)$ and $C' = C/\xi(3)$ and $\xi(3)$ denotes the scaling exponent of the 
third order SF. The vertical lines show the error bars, corresponding to the 
snapshot-to-snapshot variations. For the subsonic run, the scaling exponents in the left panel for the 
kinematic phase were measured from a single snapshot in the kinematic phase, as the number of snapshots 
available in the run after the flow is fully developed but before the magnetic field becomes dynamically 
important is very limited. 

We first  discuss the results for the subsonic flow. In the top left  panel of Fig. \ref{dim}, we see that 
$\zeta_C$ is smaller than both $\zeta_u$ and $\zeta_m$ in the kinematic phase, implying that the scalar 
structures are more intermittent than the velocity field. This is consistent with results of earlier studies for 
turbulent mixing in incompressible or weakly compressible hydrodynamical flows \citep[e.g.,][]{PS11}. 
The exponents of the mixed structures lie in between the velocity and scalar structures as 
expected. In the saturated phase (panel 1b),  $\zeta_C$, $\zeta_m$, and $\zeta_u$ are 
in the same relative order, and a comparison of the two panels shows that the velocity field 
in the saturated phase is less intermittent than in the kinetic phase, suggesting that the presence of the 
magnetic tension and pressure tends to suppress the development of intense velocity structures. 

Note that $\zeta_u$ measured here is not to be confused with the scaling exponents for the 
Els\"asser variables $z_{\pm} \equiv u \pm B/\sqrt{4 \pi}$ commonly analyzed in incompressible 
MHD turbulence studies.  We also attempted to measure the scaling of $z_{\pm}$ in our flow, and found that  
their scaling exponents are more intermittent than those ($\zeta_u$) for the velocity field itself \citep[see][]{HBD04}.  
A likely reason is that the existence of the strong magnetic structure such as the current sheets
contributes to the intermittency of $z_{\pm}$.   We point out that the difference in the scaling properties of 
the velocity field from $z_{\pm}$ suggests that the most intermittent velocity structures may have different 
locations from the strong magnetic structures, such as the current sheets. Here, for the study of mixing, 
we are mainly concerned with the velocity structures, and thus choose not to show the scalings of 
$z_{\pm}$.

In our subsonic flow with Mach 0.3, the scaling exponents, $\zeta_C$, normalized to the 3rd order 
structure function for the scalar field in the kinematic and saturated phases are very close to each other. 
However, the unnormalized exponents, $\xi_C$, for the scalar field in the two phases are different due 
to the difference in $\xi(3)$. 
This may be responsible for the different visual impressions in the scalar images shown in 
Figs.\ \ref{3dkin} and \ref{3dsat}.

As the flow Mach number changes from 0.3 to 2.4,  the velocity field in the kinematic phase becomes 
significantly more intermittent, consistent with  the earlier results of \citet{PS11} for mixing in hydrodynamical 
flows. This is due to the formation of shocks, which are intense dissipative structures and are known to 
increase the intermittency of the velocity field. On the other hand, the intermittency of the scalar structures
only changes slightly as the flow goes from subsonic to supersonic. The reason is that the existence of 
velocity shocks do not cause discontinuities in the scalar field. Since the flow and pollutant densities are 
both conserved across velocity shocks, the concentration field and the ratio of two densities, remains 
continuous (\citet{PS11}).  This explains the small changes in the scaling exponent, $\zeta_C$, of the 
scalar field, despite the significantly higher intermittency in the velocity field. In comparison to the subsonic 
case, the relative degree of intermittency is reversed with the velocity structures being slightly more 
intermittent the scalar field. 

As the magnetic field grows, the formation of shocks is expected to be suppressed, 
as the magnetic field along the shock front has the effect of opposing the converging flow, like 
the effect of the thermal pressure. Therefore, once the Alfv\'enic speed becomes comparable to 
or larger than the sound  speed, the frequency and the strength of the shocks would 
be reduced.  As a result, the velocity structures in the saturated phase become 
significantly less intermittent than in the kinematic phase.  The scaling of the concentration 
field is less affected by the growth of the magnetic strength, again because it is insensitive 
on whether velocity shocks exist. Roughly speaking, as the magnetic strength saturates 
in our $\mathcal{M}=2.4$ flow, the Alfv\'enic Mach number, which is smaller then sonic Mach 
number, would play the role of the sonic Mach number. 
The result that the velocity scaling is less intermittent than the scalar field in the saturated phase could 
be viewed as corresponding to a case with an effective Mach number significantly smaller than 2.4 
(see \citet{PS11}). 

In  summary,  the scaling of the velocity structures has 
a significant dependence on both the flow Mach number and 
the magnetic strength. The velocity intermittency increases with increasing Mach number, 
and decreases as the magnetic field becomes dynamically important and suppresses shock 
formation. On the other hand, the scaling behaviors of the scalar  field  are 
insensitive to velocity shocks, which do not produce concentration 
discontinuities, and thus show only slight changes with the Mach number and/or the magnetic growth.

Following Pan \& Scannapieco (2011), we measured the parameter $\beta_{i}$ using the structure 
function ratios, $F_{p}(r) = S_{p+1}(r)/S_{p}(r)$, at successive orders. The slope of the 
$F_{p+1}(r)/F_2(r)$ vs. $F_{p}(r)/F_1(r)$  curves at given orders, $p$, provides an estimate for 
$\beta_{i}$ (She et al.\ 2001). The values of $C'$ are then obtained by fitting the data points for 
$\zeta_{p}$ as a function of $p$ shown in the above figure. Using the measured $C'$ and the 
values of $\xi(3)$ from the third order SF's, we estimate the fractal dimension $d = 3 - \xi(3)C'$ of 
the velocity, mixed and the scalar fields. Table~\ref{datadim} summarizes the scaling exponents 
of the third order structure functions, the  parameter $\beta_{i}$ and the fractal dimension $d$ for the 
velocity, mixed and the scalar field corresponding to two ideal runs, M0.3Id2 and M2.4Id. 
The parameters given in the table for the kinematic phase are consistent with those obtained in earlier 
studies for incompressible or weakly compressible flow. The scaling exponent, $\xi_m(3)$, for the 
third-order mixed structure function is close to unity in both phases, confirming the general validity 
the scalar cascade picture.

%%%%%%%%%%%%%%%%%%%%%%%%%%%%%%%%%%%%%%%%%%%%%%%%%%
\begin{table*}
\begin{center}
\resizebox{2.0\columnwidth}{!}{%
\begin{tabular}{|c|ccc|ccc|ccc|} \hline \hline
$\mathcal{M}$ & $\xi_{u}(3)$ & $\beta_{i}^{u}$ & $d_{u}$ & $\xi_{m}(3)$ & $\beta_{i}^{m}$ & $d_{m}$ & $\xi_{C}(3)$ 
& $\beta_{i}^{C}$ & $d_{C}$ \\
 & Kin., \, Sat. & Kin., \, Sat. & Kin., \, Sat. & Kin., \, Sat. & Kin., \, Sat. & Kin., \, Sat. & Kin., \, Sat. & Kin., \, Sat. 
 & Kin., \, Sat. \\ \hline \hline 
0.3 & 0.98 \, 1.40 & 0.88 \, 0.88 & 1.04 \, 1.00 & 0.96 \, 0.96 & 0.74 \, 0.81 & 2.08 \, 1.84 & 0.87 \, 0.65 & 0.68 \, 0.64 
& 2.15 \, 2.41 \\ \hline 
2.4 & 1.15 \, 1.33 & 0.70 \, 0.70 & 1.50 \, 1.94 & 0.95 \, 0.96 & 0.75 \, 0.74 & 1.58 \, 2.03 & 0.78 \, 0.71 & 0.63 \, 0.63 
& 2.25 \, 2.47 \\ \hline \hline
\end{tabular}
}
\end{center}
\caption{Scaling exponent $\xi(3)$ of the third order SF's, intermittency parameter $\beta_{i}$ and the fractal 
dimension $d$ of the most intense structures for M0.3Id2 and M2.4Id runs.}
\label{datadim}
\end{table*}
%%%%%%%%%%%%%%%%%%%%%%%%%%%%%%%%%%%%%%%%%%%%%%%%%%%%%%%%%%

We find that for each type of structures, the velocity, mixed or scalar, there exists a narrow 
range of $\beta_{i}$ around the values listed in Table 2 that can fit $\zeta(p)$ will 
as a function of $p$.  In such a range of $\beta_{i}$, one can tune the parameter $C'$ to obtain satisfactory 
fits to $\zeta(p)$. It turns out that, for the velocity structures in the subsonic run, especially in the saturated 
phase, the best-fit $C'$ and hence $d$ have a very sensitive dependence on the choice of $\beta_{i}$. 
For example, if we set $\beta_{i}^{u}$ in the saturated phase to 0.86, the best-fit $C'_{u}$ would be 1.05 
and thus $d_{u}=1.53$. On the other hand, if we increase $\beta_{i}^{u}$ slightly to 0.90, we obtained $C'_{u}=2$ 
and $d_{u}=0.2$. Such a sensitive dependence suggests that $d_{u}$ cannot be measured reliably unless 
$\beta_{i}^{u}$ can be determined at an accuracy level $\ll 1\%$. 
Such an accuracy level is not possible in our simulations, the definite determination of the 
dimension of the strong velocity structures in the saturated phase requires high-resolution simulations 
that allow a broad inertial range. For the mixed and scalar structures and all the three kinds of structures in 
the supersonic run, the dependence of $C'$ or $d$ on the selected value of $\beta_{i}$ is much weaker than 
in the velocity case, and the measured values of $d$ are thus reliable at least qualitatively.

For M0.3Id2 run, the fractal dimension $d_{C}$ for the scalars changes slightly from $2.15$ in the 
kinematic phase to $2.41$ in the saturated phase, suggesting that the strong scalar structures continue 
to be sheet-like as the magnetic field grows to saturation. A very similar change in the scalar dimension 
is observed in the supersonic case. The slightly larger $d_{C}$ in the saturated phase 
appears to be consistent with the morphological differences in Figs.\ \ref{3dkin} and \ref{3dsat}, and is
likely arises because the Lorentz force resists compression, and thus  cliffs become thicker.
The dimensionality of the strong scalar structures may be understood by an analysis of the strain 
tensor.  For example, if the scalar field is compressed  in two directions, then the 
scalar structures would tend to be filamentary. On the other hand, our finding of sheet-like scalar 
structures in both phases implies that  the strain tensor compresses the scalar field in one direction 
and extends it in the other two directions. We will perform a more detailed study of the strain tensor and 
its relation to the magnetic and concentration field structures in a future work. 

\subsection{Scalar Dissipation and Mixing Timescale}

The small-scale dynamo generates magnetic fields that are random in both space and time as evident 
from Figs.~\ref{3dkin} and \ref{3dsat}. At the same time, turbulent eddies also stretch the scalar 
concentration field forming similar random small-scale structures. Here, we explore how the turbulent 
mixing varies as a function of the magnetic field and how does the mixing time scale depends on the 
Prandtl and Peclet numbers and the Mach number of the flow. To study this effect, we compute the mean 
of the scalar dissipation $\langle\tilde{\rho}(\nabla C)^2\rangle$ conditioned on the magnetic field strength, 
where again $\tilde{\rho}\equiv\rho/\overline{\rho}$ is the ratio of the density $\rho$ to the average flow 
density, $\overline{\rho}$. As is clear from eq.\ (\ref{eq:Cvariance}), scalar dissipation is strongest in 
regions in which $\tilde{\rho}(\nabla C)^2$ is large, which reduces to $(\nabla C)^2$ in the incompressible 
limit. As  large scalar gradients would therefore imply efficient mixing, it is important to study how 
$\tilde{\rho}(\nabla C)^2$ varies in regions with different  magnetic field strengths, as produced by a 
small-scale dynamo. 
%%%%%%%%%%%%%%%%%%%%%%%%%%%%%%%%%%%%%%%%%%%%%%%%%%%%
\begin{table*}
\begin{center}
\resizebox{2.0\columnwidth}{!}{%
\begin{tabular}{|c|c|c|c|c|c|c|c|} \hline \hline 
Simulation 
& $\langle\tilde{\rho}(\nabla C)^2\rangle$ & $\langle\tilde{\rho}(\nabla C)^2\rangle$ 
& $\langle\tilde{\rho}(\nabla C)^2\rangle$ & $\langle\tilde{\rho}(\nabla C)^2\rangle$ 
& $\langle\tilde{\rho}(\nabla C)^2\rangle$ & $\langle\tilde{\rho}(\nabla C)^2\rangle$ 
& $ 100 \times \langle\tilde{\rho} C^2\rangle$\\ 
Run & &  $\frac{B}{B_{\rm rms}} \leq 0.5$ & $0.5 <\frac{B}{B_{\rm rms}} \leq 1.0$ & $1 < \frac{B}{B_{\rm rms}} \leq 2$ 
& $2 \frac{B}{B_{\rm rms}} \leq 3$ & $3 <\frac{B}{B_{\rm rms}} \leq 5$ & \\ 
&  Kin., \,\, Sat. & Kin., \,\, Sat. & Kin., \,\, Sat. & Kin., \,\, Sat. & Kin., \,\, Sat., & Kin., \,\, Sat. & Kin., \,\, Sat. \\ \hline \hline
% Ideal run 
M0.3Id2 &  107, \, 197 & 107, \,  175 & 108, \, 201 & 128, \, 217 & 167, \, 206 & 207, \, 177 &  0.607, \,     0.988 \\ \hline
% Pm=1, Beta=10^4 run
M0.3Pm1a &  19.6, \, 20.6  & 15.2, \, 18.1 & 19.9, \, 21.2 & 24.8, \, 22.8 & 28.3, \, 21.3 & 31.8, \, 17.2 & 0.608, \,      0.862 \\ \hline
% Pm=2, Beta=10^4 run
M0.3Pm2  &  8.85, \, 9.35 & 7.34, \, 8.07 & 8.98, \, 9.40 & 10.7, \, 10.6 & 12.0, \, 10.2 & 13.2, \, 10.2  &   0.587, \,      0.742 \\ \hline
% Pm=4, Beta=10^4
M0.3Pm4 & 4.87, \, 5.07 & 4.07, \, 4.70 & 4.94, \, 5.26 & 5.90, \, 5.46 & 6.70, \, 4.88 & 7.18, \, 4.60  & 0.543, \,    0.675\\ \hline 
% Pm =1, M = 1.1 
M1.1Pm1 &  21.6, \, 24.1 & 14.2, \, 18.9 & 21.7, \, 24.3 & 30.9, \, 28.3 & 41.2, \, 29.9 & 50.8, \, 32.7 & 0.735, \,      1.00\,\,  \\ \hline  
% M=2.4, Ideal
M2.4Id &  178, \, 281 & 74.0,\,\,\,137 & 167,\,\,\, 243 & 330,\,\,\, 421 & 622,\,\,\, 774 & 1184,\,\, \,1350 &  0.866, \,\,\,       1.47\,\,   \\ \hline
%M2.4Id & 47.5,\,137 & 102,\,\, 243 & 186,\,\, 421 & 324,\,\, 774 & 498,\,\, 1350 & 95.3, \, 281\\ \hline
% Pm=1, M=2.3 
M2.3Pm1 &  27.2, \,\,\,\,\,--\,\,\, & 9.64, \,\,\,\,-- \,\,\,& 26.2,\,\,\,\,\,--\,\,\, & 55.14,\,\,\,\,\,-- \,\,\,& 105,\,\,\,\,\,-- & 174,\,\,\,\,\,\,\,--   & 0.910,\,\,\,\,\,\,\,\,\,-- \, \\ \hline \hline
\end{tabular}
}
\end{center}
\caption{Mean values of $\langle\tilde{\rho}(\nabla C)^2\rangle$ for the different simulation runs in both 
the kinematic and saturated phases. The overall values of the scalar dissipation are tabulated in the 
second column while the values conditioned on the magnetic field strength are presented in the 3rd to 
the 7th columns. For reference, $\langle\tilde{\rho} C^2\rangle$ is presented in the 8th column}.
For M2.3Pm1, we only show the values in the kinematic phase. Note that while all the 
values in this table can be rescaled by a single arbitrary number, corresponding to the strength of the 
scalar driving, the relative values can be accurately compared across runs, across magnetic field ranges, 
and between the kinematic and saturated phases.
\label{hist1}
\end{table*}
%%%%%%%%%%%%%%%%%%%%%%%%%%%%%%%%%%%%%%%%%%%%%%%%%%%%%%%%%

We show the results obtained from the various runs in Table~\ref{hist1}. For subsonic turbulence, 
$\langle\tilde{\rho}(\nabla C)^2\rangle$ increases with increasing field strength in the kinematic phase.
Because this occurs in a phase in which the magnetic field is unimportant, this implies that magnetic field
amplification and scalar gradients are both increased by the same overall properties of the local flow, 
in particular, the velocity gradient. In fact, a simultaneous increase of both these quantities is expected in 
regions in which the flow is contracting relative to the direction of the local magnetic field, as this would both 
increase the field amplitude by flux freezing, and simultaneously increase the scalar gradient in the direction 
of the contraction. In the saturated phase, on the other hand, $\langle\tilde{\rho}(\nabla C)^2\rangle$ 
increases with increasing field strength at low magnetic field values, and then decreases with increasing 
magnetic field strength, hinting at Lorentz forces inhibiting the buildup of large scalar gradients.  

%%%%%%%%%%%%%%%%%%%%%%%%%%%%%%%%%%%%%%%%%%%%%%%%%%%%%%
\begin{figure}[h]
%\begin{center}
\hspace{-0.14in}
\includegraphics[width=1.07\columnwidth]{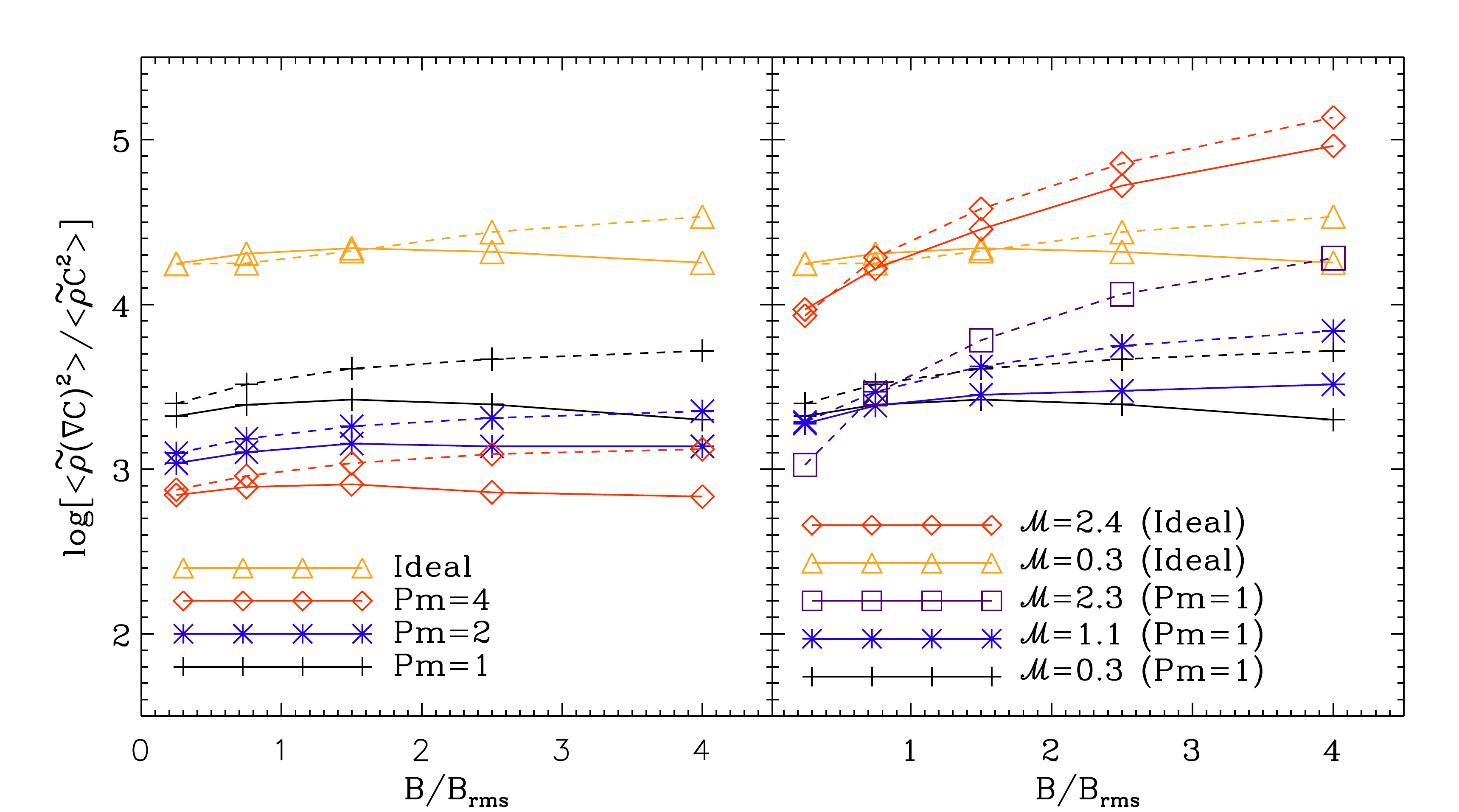} 
%\end{center}
\vspace{-.1in}
\caption{The values of $\langle\tilde{\rho}(\nabla C)^2\rangle/\langle\tilde{\rho} C^2\rangle$ for the different 
${\rm Pm}$ runs in the left panel and for the different Mach number runs in the right panel. Dashed lines 
denote the kinematic phase while the solid lines are for the saturated phase. }
\label{tabfig}
\vspace{0.1in}
\end{figure}
%%%%%%%%%%%%%%%%%%%%%%%%%%%%%%%%%%%%%%%%%%%%%%%%%%%%%%
  
We observe these trends to hold irrespective of the magnetic Prandtl numbers, although 
the overall magnitude of $\langle\tilde{\rho}(\nabla C)^2\rangle$ decreases strongly with Pm, as 
expected because increasing $\kappa$ will increase the length scale of typical scalar structures. 
On the other hand, for $\mathcal{M}=1.1$ and $2.4$ the scalar dissipation, $\langle\tilde{\rho}(\nabla C)^2\rangle$ 
continues to increase monotonically with the magnetic field strength in both the kinematic and saturated phases.
This is most likely due to the fact that the ratio of the magnetic to kinetic energies, $E_{\rm m}/E_{\rm k}$ is 
still low even in the saturated phase compared to the subsonic runs. The monotonic increase of 
$\langle\tilde{\rho}(\nabla C)^2\rangle$ is the strongest for M2.4Id run among all the runs reported here. 
Since we could only follow the M2.3Pm1 run till the linear growth phase, we only show the corresponding 
values for the kinematic phase. For reference,  in the last column in Table~\ref{hist1}, we give the overall 
density-weighted concentration variance $\langle\tilde{\rho} C^2\rangle$, not conditioned on the magnetic 
field strength.  In Figure~\ref{tabfig}, we show the conditional dissipation rate normalized to the variance 
$\langle\tilde{\rho} C^2\rangle$, as a function of the magnetic strength. Clearly, the normalization provides 
a better measure for the mixing efficiency. In general, the normalized conditional dissipation in the saturated 
phase is lower than in the kinematic phase, again supporting the argument that strong magnetic fields tend 
to slow down the buildup of the scalar gradients.

%%%%%%%%%%%%%%%%%%%%%%%%%%%%%%%%%%%%%%%%%%%%%%%%%%%%%%
\begin{figure}[h]
%\begin{center}
\hspace{-.25in}
\includegraphics[width=1.12\columnwidth]{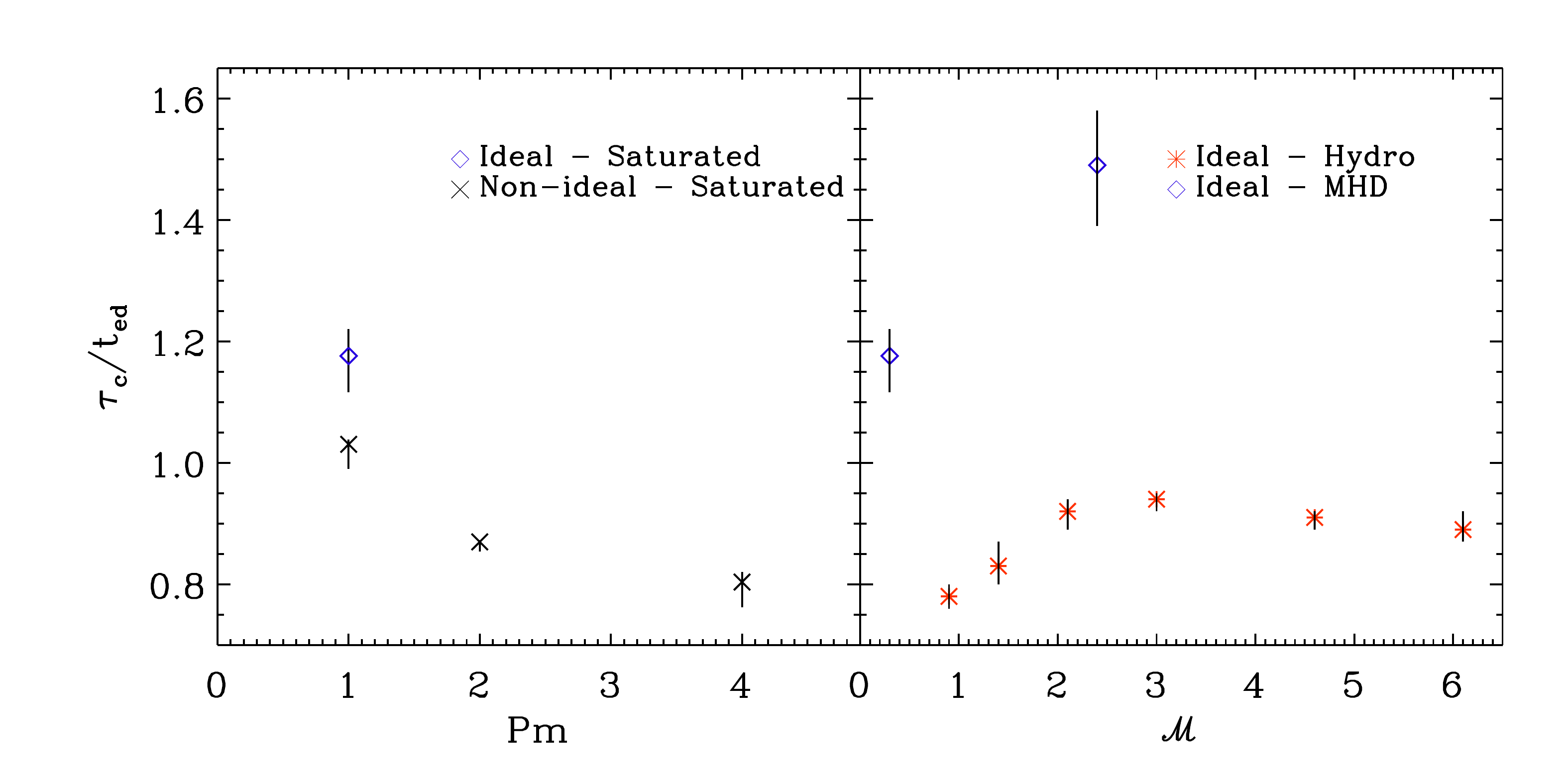}
%\end{center}
\caption{Dependence of the mixing time scale $\tau_{c}$ on flow properties. In the left panel, labeled 
as a function of Prandtl number,  we see that the mixing timescale in the saturated phase decreases 
as the magnetic resistivity is held fixed, and viscosity and scalar diffusivity are increased simultaneously.   
In the right panel, the mixing time scale in ideal runs with varying  Mach number are plotted in the saturated 
phase (diamonds) and  from purely hydrodynamical simulations \citep[asterisks,][]{PS10} In both panels 
the solid  vertical lines denote error bars.
\label{timescale}}
\end{figure}
%%%%%%%%%%%%%%%%%%%%%%%%%%%%%%%%%%%%%%%%%%%%%%%%%%%%%%

Finally, we computed  the overall rate of mixing as a function of Prandtl number and Mach number.
In the presence of a physical scalar diffusivity, the mixing time scale $\tau_{\rm c}$ is defined as 
\EQ
\tau_c = \langle\tilde{\rho} C^2\rangle/ \overline{\epsilon}_c = 
\langle\tilde{\rho} C^2\rangle/ 2\langle\tilde{\rho}\kappa (\nabla C)^2\rangle,
\EN
where $\overline{\epsilon}_{c} = 2\langle\tilde{\rho}\kappa (\nabla C)^2\rangle$ is the scalar dissipation rate. 
In ideal MHD, the above definition no longer holds since the scalar diffusivity is of numerical origin. 
However, as can be seen from eq.\ (\ref{eq:Cvariance}), in the statistically homogeneous case,
one can still define a mixing time scale by assuming a balance between the scalar dissipation rate and 
the scalar source term when scalar fluctuations reach a statistically stationary state \citep{PS10}. This then 
leads to a dissipation time scale, $\tau_c = \langle\tilde{\rho} C^2\rangle/\overline{\epsilon}_{s} = 
\langle\tilde{\rho} C^2\rangle/\langle\tilde{\rho} T C\rangle$ where 
$\overline{\epsilon}_{s}=\langle\tilde{\rho} T C\rangle$ denotes the source term for scalar fluctuations. 
To check the validity of this assumption we computed $\tau_{c}$ by both the methods for one of our 
non-ideal runs (M0.3Pm1a) and find the two time scales to be similar got  $t  \geq 5\,t_{\rm ed}$. 

In Figure~\ref{timescale}, we show the dependence of the mixing time scale $\tau_{\rm c}$ normalized to 
the eddy turnover time on the magnetic Prandtl number ${\rm Pm}$ (left panel) and Mach number (right panel) 
when the magnetic field has reached saturation. The mixing time scale shows a weak dependence on the 
magnetic Prandtl number with the timescale decreasing with increasing Prandtl number.
As discussed above, we increase the viscosity  and scalar diffusivity simultaneously, in our simulations to 
obtain ${\rm Pm} > 1$. This results in a drop of Reynolds and Peclet numbers from ${\rm Re=Pe}$ from 
$1250$ for ${\rm Pm}=1$ to $\approx 312$ for ${\rm Pm}=4$. In other words, the fluid tends to become more 
viscous, and, likely more importantly, $\kappa$ becomes higher, making scalar diffusion more effective and 
reducing the mixing time scale.  Note that we have assumed that ${\rm Pm} \approx O(1)$ for the ideal run. 

In the presence of magnetic fields, the mixing time scale increases with increasing Mach number as can 
be seen from the right panel in Fig.~\ref{timescale}. As first measured in \citet{PS10} and also shown in this 
figure, this trend holds even for pure hydrodynamical runs where the time scale increases with the increase 
in Mach number from $\mathcal{M}=0.9$ to $3.0$ and thereafter remains constant. Moreover, for a given 
Mach number, the time scale in the MHD case is higher than its hydrodynamic counterpart, which 
suggests that the presence of magnetic fields slows down the mixing process. This is consistent with 
a physical picture in which the magnetic tension resists the straining of the flow, thereby leading to 
a decrease in the rate of stretching. 

\section{Summary and Conclusions}
\label{summ}

Magnetic fields are ubiquitous in a wide variety of astrophysical objects ranging from molecular 
clouds to galaxies and galaxy clusters, as they are both amplified and maintained by turbulence. 
Such random motions are driven by many sources ranging from  supernovae in the ISM 
\citep{deAB05, Hill+12, Gent+13} to structure formation and active galactic nuclei in galaxy clusters 
\citep{NB99, Ryu+08, Ryu+12}. They are also responsible for the turbulent mixing of elements, which 
is essential for metal enrichment in the IGM, the pollution of the primordial gas by the first generation 
of stars, and our interpretation of a variety of observations such as the metallicity dispersion in open 
clusters, the abundance scatter in the ISM, and the cluster to cluster metallicity scatter.  Here we have 
carried out a systematic investigation into the impact of magnetic fields on turbulent mixing by performing 
three-dimensional simulations of forced turbulence in a box including magnetic fields and the injection 
of pollutants at regular intervals. With this setup, we have explored two different parameter regimes 
consisting of simulations with different magnetic Prandtl numbers, ${\rm Pm} = 1, 2$ and $4$ for 
$\mathcal{M}=0.3,$ and runs with varying Mach numbers, $\mathcal{M}=0.3, 1.1$ and $2.3$ for 
${\rm Pm}=1$. In addition, we have also performed ideal simulations to compare with our non-ideal 
runs at $\mathcal{M}=0.3$ and at $2.4$. 

Small-scale dynamo action initially amplifies the magnetic fields exponentially, then linearly, and 
eventually saturates when energy equipartition is reached on scales close to the driving scale of 
turbulence. The saturated phase of the small-scale dynamo is crucial to our studies since, in this phase, 
the field is strong enough to back-react on the flow and affect its mixing properties. Due to physical 
viscosities and diffusivities in our non-ideal simulations, the magnetic and the scalar field structures 
appear to be thicker than their ideal MHD counterparts (see Figs.~\ref{3dkin} and \ref{3dsat}). Nevertheless, 
in both cases we find that the magnetic field is strong only in certain regions of the simulation, and that the 
coherence length of the field increases in the saturated phase. The scalar concentration field also shows 
sharp contrasts in the kinematic phase, resulting from random stretching and shearing by turbulent eddies. 
Since most of our simulations begin with $\beta_{\rm in}=10^{4}$, we do not have a prominent kinematic 
growth phase to compute the dependence of the growth rates on the magnetic Prandtl number and the Mach 
number of the flow. However, we find a decrease in the growth rate of the magnetic field with increasing 
Mach numbers when comparing the M0.3Id2 and the M2.4Id runs (see inset figure in  Fig~\ref{tseries1}).

In the kinematic phase, the velocity power spectrum for  $\mathcal{M}=0.3$ has a $k^{-5/3}$ slope and the 
magnetic power spectrum has a slightly flatter scaling compared to the $k^{3/2}$ Kasantsev's scaling. 
As evident from panel 1(b) in Fig.~\ref{spectra}, most of the field amplification in this phase is  driven by 
turbulent eddies on scales close to the dissipation scale, due to their shorter eddy turnover times, and this 
results in the magnetic energy peaking on small scales. However, like the velocity spectra, the scalar power 
spectrum shows a $k^{-5/3}$ scaling in the kinematic phase for the subsonic run. 

In the saturated phase, the velocity spectrum becomes steeper than $k^{-5/3}$ due to the faster transfer of 
kinetic energy to magnetic energy at small scales. As a response to the steepening of the velocity spectrum, 
the scalar spectrum becomes flatter, $~ k^{-4/3},$ because a decrease in the velocity power at smaller scales 
implies an increase in the scalar fluctuations. For supersonic, hydrodynamic turbulence, it is well known that 
the velocity spectrum steepens from $k^{-5/3}$ to $k^{-2}$ due  to the dominance of shocks. In the saturated 
phase, we find that the velocity spectrum for $\mathcal{M}=2.4$  becomes marginally steeper than $k^{-2}$ 
scaling due to magnetic back reactions. In this case, the scalar spectra shows a $\approx k^{-1.42}$ scaling 
compared to the $\approx k^{-4/3}$ scaling for the subsonic and transonic runs in the saturated phase. 

We also studied the structure functions (SF's) for the velocity and the concentration fields 
for our ideal MHD runs at $\mathcal{M}=0.3, 2.4$ due to their longer inertial range compared to the 
other non-ideal runs. For the subsonic case, the longitudinal velocity and the scalar SF's 
at the second order show an $\approx l^{2/3}$ scaling in the kinematic phase as expected. 
As the magnetic field grows to  saturation, the scaling steepens to $l^{1.05}$ for the velocity, 
while for the concentration it flattens to $l^{1/2}$.  
In the supersonic case, the second order SF's for the velocity and scalar fields
scale as $\approx l^{0.85}$ and $\approx l^{0.56}$ in the kinematic phase, and change 
to $\approx l^{0.92}$ and $\approx l^{0.52}$, respectively, in the saturated phase. 
In both cases, the scaling exponents for the concentration and velocity fields agree 
remarkably well with a relation predicted by the OC picture. 

Quantitative analysis of higher order SF's show that, for subsonic turbulence, the scalar concentration 
field is more intermittent than the velocity field in both the kinematic and the saturated phases. As the 
magnetic filed increases, the velocity becomes slightly less intermittent, while the scalar intermittency 
remains roughly unchanged. At a flow Mach number of  2.4, the existence of shocks greatly increases 
the intermittency of the velocity field in the kinematic phase, making it more intermittent than the scalar 
field. Interestingly, with the growth of the magnetic field strength to saturation, the formation of shocks is 
suppressed by the magnetic pressure, and this significantly reduces the intermittency of the velocity field, 
bringing it back to a level below the scalar intermittency. Our results show that, unlike the velocity 
structures whose intermittency depends on the existence and strength of shocks, the scalings of high-order 
scalar structures appear to be insensitive to the flow Mach number or the magnetic growth. Independent 
of the Mach number of the flow, the strong scalar structures are sheet-like with a fractal dimension 
$d_{c} \approx 2$ in both phases of magnetic field evolution. This implies that the effect of an evolving 
magnetic field on the topology of the strong scalar structures is slight.

We find that the scalar dissipation becomes less efficient as the magnetic field grows from the kinematic 
phase to saturation.  As shown in  Fig.~\ref{tabfig},  when normalized to the scalar variance, the dissipation 
rate is generally smaller in the saturated phase for both the subsonic and supersonic runs. We analyzed the 
dissipation rate conditioned on the magnetic field strength and showed that in the saturated phase of the 
subsonic run the dissipation becomes slower at sufficiently large $B$, indicating that mixing is hindered 
in regions with strong magnetic field. These all suggest that the Lorentz force resists  the straining of the 
flow and hence suppressing the buildup of large scalar gradients, leading to slower mixing when the magnetic 
field reaches saturation. We observe this trend to hold independent of the magnetic Prandtl number. Unlike 
the subsonic run, in the saturated phase of the  supersonic case, the scalar dissipation conditioned on $B$ 
keeps increasing monotonically with increasing with the magnetic field strength, which can be attributed to 
the fact that the $E_{\rm m}/E_{\rm k}$ ratio is still low even in the saturated phase. 

Analysis of the mixing time scale, $\tau_{c},$ reveals a weak dependence on the magnetic Prandtl number, 
with $\tau_{c}$ decreasing with increasing ${\rm Pm}$. In our simulations, we have achieved ${\rm Pm}>1$ by 
decreasing the fluid Reynolds and Peclet numbers simultaneously, making scalar
diffusion more effective and reducing the mixing time scale. Most astrophysical systems like galaxies and 
galaxy clusters have ${\rm Pm}\gg 1$ and it would be interesting to see how $\tau_{c}$ changes 
in this regime. Computationally, attaining such large ${\rm Pm}$ with a high ${\rm Re}$ is strongly limited 
by the resolution of the simulations, but the case with low ${\rm Re}$, although not corresponding directly to  any  
astrophysical system, may nevertheless be worthy of a systematic investigation. 
The mixing time scale increases with increasing Mach number in our simulations. However, more simulations 
at $\mathcal{M}>2.4$ are needed  to draw definitive conclusions. Earlier simulations of 
\citep{PS10} also find an increase in mixing time scale with increasing Mach numbers up to $\mathcal{M} = 3$, 
beyond which it remains constant. 

Our results also show a higher value of $\tau_{c}$ in the MHD case compared to the hydro case at a given 
Mach number. This is due to the fact that once the magnetic field reaches saturation, magnetic tension will 
resist the straining of the flow, thereby decreasing the rate of stretching and inhibiting large scalar gradients. 

In this paper, we have concentrated on simulations where the Schmidt number ${\rm Sc}=1$. It would be 
interesting to explore the parameter regime where ${\rm Sc}$ is different from unity. For example, ${\rm Sc}>1$ 
would imply that the scalar dissipation scale lies outside the viscous dissipation scale while for ${\rm Sc}<1$, 
the scalar dissipation scale would lie inside the viscous dissipation scale. 

Although pollutants such as metals constitute only a negligible fraction of the cosmic matter budget, they play 
a crucial role in constraining star formation, tracing feedback from massive stars and supernovae. They also 
play a dominant role in the cooling of gas thereby affecting large-scale structure formation on a wide range of 
scales. Our present work provides a first step toward understanding how such processes are influenced by 
magnetic fields.

\acknowledgments

SS \& ES were supported by the National Science Foundation under grant AST11-03608 and NASA theory 
grant NNX09AD106. LP acknowledges support from the Clay postdoctoral fellowship at Harvard-Smithsonian 
Center for Astrophysics. The authors would also like to acknowledge the Advanced Computing Center at 
Arizona State University (URL: http://a2c2.asu.edu/), the Texas Advanced Computing Center (TACC) at The 
University of Texas at Austin (URL: http://www.tacc.utexas.edu), and  the Extreme Science and Engineering 
Discovery Environment (XSEDE) for providing HPC resources via grant TG-AST130021 that have contributed 
to the research results reported within this paper. We would like to thank Robert Fisher, Dongwook Lee and 
Dean Townsley for helpful discussions. The FLASH code is developed in part by the DOE-supported 
Alliances Center for Astrophysical Thermonuclear Flashes (ASC) at the University of Chicago. 

\newpage
%\bibliographystyle{apj}
%\bibliography{msbib}

\end{document}